\documentclass[twocolumn]{autart}%
\usepackage{graphicx}
\usepackage{epsfig}
\usepackage{color}
\usepackage{enumerate}
\usepackage{cite}
\usepackage{graphicx}
\usepackage{subcaption}
\usepackage{romannum}
\usepackage{epstopdf}
\usepackage{balance}
\usepackage{bigints}
\usepackage{amssymb}
\usepackage{multirow}
\usepackage{amsmath}
\usepackage{hyperref}
\usepackage[utf8x]{inputenc}
\usepackage{graphicx}
\usepackage{epsfig}
\usepackage{mathptmx}
\usepackage{times}
\usepackage{amsmath}
\usepackage{amssymb}
\usepackage[section]{placeins}
\usepackage{placeins}
\usepackage{amsmath}
\usepackage{multirow}
\usepackage{graphicx}
\usepackage[normalem]{ulem}
\usepackage{hyperref}
\usepackage{subcaption}
\usepackage{algorithm,tabularx}
\usepackage{algpseudocode}
\usepackage{amsfonts}
\usepackage{cases}
\usepackage{romannum}
\usepackage{epstopdf}
\usepackage{textcomp}
\usepackage{soul}
\usepackage{mathtools}
\usepackage{subcaption}
\usepackage{textcomp}
\usepackage{color}
\usepackage{url}%

\providecommand{\U}[1]{\protect\rule{.1in}{.1in}}
\providecommand{\U}[1]{\protect\rule{.1in}{.1in}}
\renewcommand{\thefootnote}{\roman{footnote}}

\usepackage[section]{placeins}
\makeatletter
\AtBeginDocument{%
  \expandafter\renewcommand\expandafter\subsection\expandafter{%
    \expandafter\@fb@secFB\subsection
  }%
}
\makeatother

\begin{document}
\begin{frontmatter}
% \title{{\LARGE \textbf{Optimal Composition of Heterogeneous Multi-Agent Teams for
% Coverage Problems with Performance Bound Guarantees}}}
% \author{Chuangchuang Sun, Shirantha Welikala and Christos G. Cassandras
% \thanks{$^{\star}$Supported in part by NSF under grants ECCS-1509084,
% DMS-1664644, CNS-1645681, by AFOSR under grant FA9550-19-1-0158, by ARPA-E's
% NEXTCAR program under grant DE-AR0000796 and by the MathWorks.}
% \thanks{Chuangchuang Sun is with Department of Aeronautics and Astronautics,
% Massachusetts Institute of Technology, Cambridge, MA 02139, USA; Work
% performed while at Boston University. Email: \texttt{{\small ccsun1@mit.edu.}}
% Shirantha Welikala and Christos G. Cassandras are with the Division of Systems
% Engineering and Center for Information and Systems Engineering, Boston
% University, Brookline, MA 02446, USA. Emails: \{\texttt{{\small shiran27,
% cgc\}@bu.edu.}}} }
% \maketitle
\title{Optimal Composition of Heterogeneous Multi-Agent Teams for
Coverage Problems with Performance Bound Guarantees\thanksref{footnoteinfo}}
\thanks[footnoteinfo]{This paper was not presented at any
conferences. For C. Sun, this work was performed while at Boston University. This work is supported in part by NSF under grants ECCS-1509084,
DMS-1664644, CNS-1645681, by AFOSR under grant FA9550-19-1-0158, by ARPA-E's NEXTCAR program under grant DE-AR0000796 and by the MathWorks.}
\author[A]{Chuangchuang Sun}\ead{ccsun1@mit.edu},    % Add the
\author[B]{Shirantha Welikala and Christos G. Cassandras}\ead{\{shiran27, cgc\}@bu.edu.}% e-mail address
%\author[Baiae]{Publius Maro Vergilius}\ead{vergilius@culture.ir}  % (ead) as shown
\address[A]{Department of Aeronautics and Astronautics,
Massachusetts Institute of Technology, Cambridge, MA 02139, USA.}  % Please supply
\address[B]{Division of Systems Engineering and Center for Information and Systems Engineering, Boston University, Brookline, MA 02446, USA.}             % full addresses
%\address[Baiae]{The White House, Baiae}        % here.
% \author{Chuangchuang Sun, Shirantha Welikala and Christos G. Cassandras
% \thanks{$^{\star}$Supported in part by NSF under grants ECCS-1509084,
% DMS-1664644, CNS-1645681, by AFOSR under grant FA9550-19-1-0158, by ARPA-E's
% NEXTCAR program under grant DE-AR0000796 and by the MathWorks.}
% \thanks{Chuangchuang Sun is with Department of Aeronautics and Astronautics,
% Massachusetts Institute of Technology, Cambridge, MA 02139, USA; Work
% performed while at Boston University. Email: \texttt{{\small ccsun1@mit.edu.}}
% Shirantha Welikala and Christos G. Cassandras are with the Division of Systems
% Engineering and Center for Information and Systems Engineering, Boston
% University, Brookline, MA 02446, USA. Emails: \{\texttt{{\small shiran27,
% cgc\}@bu.edu.}}} }
\maketitle
 %\IEEEpeerreviewmaketitle
\begin{abstract}
We consider the problem of determining the optimal composition of a heterogeneous multi-agent team for coverage problems by including costs associated with different agents and subject to an upper bound on the maximal allowable number of agents. We formulate a resource allocation problem without introducing additional non-convexities to the original problem. We develop a distributed Projected Gradient Ascent (PGA) algorithm to solve the optimal team composition problem. To deal with non-convexity, we initialize the algorithm using a greedy method and exploit the submodularity and curvature properties of the coverage objective function to derive novel tighter performance bound guarantees on the optimization problem solution. Numerical examples are included to validate the effectiveness of this approach in diverse mission space configurations and different heterogeneous multi-agent collections. Comparative results obtained using a commercial mixed-integer nonlinear programming problem solver demonstrate both the accuracy and computational efficiency of the distributed PGA algorithm.
\end{abstract}
\end{frontmatter}
%The paper headers
%\markboth{Journal of \LaTeX\ Class Files,~Vol.~14, No.~8, August~2015}{Shell \MakeLowercase{\textit{et al.}}: Bare Demo of IEEEtran.cls for IEEE Journals}

%%%%%%%%%%%%%%%%%%%%%%%%%%%%%%%%%%%%%%%%%%%%%%%%%%%%%%%%%%%%%%%%%%%%%%%%%%%%%%%%

%\thispagestyle{empty} \pagestyle{empty}

%%
%\begin{keywords}
%distributed optimization, projected subgradient algorithm, convex mixed-integer program
%\end{keywords}

%%%%%%%%%%%%%%%%%%%%%%%%%%%%%%%%%%%%%%%%%%%%%%%%%%%%%%%%%%%%%%%%%%%%%%%%%%%%%%%%

%\vspace{-0.25cm}

\section{Introduction}

\pagenumbering{arabic} Cooperative multi-agent systems are pervasive in a
number of applications, including but not limited to,
surveillance~\cite{castanedo2010data,vallejo2011multi}, search and rescue
missions~\cite{luo2011multi},
consensus \cite{Zheng2011,Zheng2018} and
agriculture~\cite{balmann2001modeling}. One of the most basic tasks such a
system can perform that has seen a wide range of applications is
\emph{coverage}. The fundamental multi-agent optimal coverage problem has been
extensively studied in the literature,
e.g.,~\cite{cassandras2005sensor,meguerdichian2001coverage,caicedo2008coverage,caicedo2008performing,breitenmoser2010voronoi}%
. In this problem, agents are deployed to \textquotedblleft
cover\textquotedblright\ as much of a given mission space as possible in the
sense that the team aims to optimally jointly detect events of interest (e.g.,
data sources) that may randomly occur anywhere in this space. The coverage
performance is measured by an appropriate metric, which is normally defined as
the joint event detection probability. The optimal coverage problem is
particularly challenging due to the generally non-convex nature of this metric
and the non-convexity of the mission space itself due to the presence of
obstacles which act as constraints on the feasible agent locations that
constitute a solution to the problem.

Thus far, the analysis of the optimal coverage problem has been carried out
based on the assumption that there exists a fixed number $N$ of agents to be
deployed. However, this number is often limited by cost constraints, leading
to a natural trade-off between coverage performance (which is normally
monotonically increasing in $N$) and total system cost. In such a setting, an
additional aspect of the problem is that of managing a set of
\emph{heterogeneous} agents: when agents fall into different classes
characterized by different properties such as sensing capacity, range,
attenuation rate, and cost, then the problem becomes one of determining the
\emph{optimal cooperative team composition} in terms of the number of agents
selected from each class so as to optimize an appropriate metric capturing the
performance-cost trade-off. Clearly, it is possible that a certain team
composition can achieve the same coverage performance as another, but with a
lower cost due to the heterogeneity of agents. The purpose of this paper is to
address the optimal coverage problem in the presence of heterogeneous agents
under cost constraints.

As mentioned above, the optimal coverage problem is already challenging due to
its non-convex nature. Heuristic algorithms (e.g., genetic
algorithms~\cite{Davis1996}), are often used and may lead to empirically
near-global optimality, but they are prohibitively inefficient for on-line
use. On the other hand, on-line algorithms sacrifice potential optimality to
achieve efficiency; this includes distributed gradient-based
algorithms~\cite{Zhong2011,cassandras2005sensor,gusrialdi2011distributed} and
Voronoi-partition-based
algorithms~\cite{breitenmoser2010voronoi,cortes2004coverage,gusrialdi2008voronoi}
which lead to generally locally optimal solutions. Methods for efficiently
escaping such local optima using a \textquotedblleft boosting
function\textquotedblright\ approach were proposed
in~\cite{Sun2014,Welikala2019P1}, while a decentralized control law
in~\cite{Schwager2008} seeks a combination of optimal coverage and exploration
of the area of interest.

A parallel effort to deal with the difficulty of finding a globally optimal
solution for the basic coverage problem is by exploiting the
\emph{submodularity} properties of the coverage performance functions used
(e.g., the joint event detection probability). This is accomplished in
\cite{Sun2019} by using a greedy algorithm to initialize the state of the
system (i.e., the locations of the agents), followed by a conventional
gradient ascent technique to obtain an improved (still locally optimal)
solution. Due to submodularity, the ratio $f^{G}/f^{\ast}$, where $f^{G}$ and
$f^{\ast}$ correspond to the objective function values under a greedy solution
and the globally optimal solution respectively, has a lower bound $L\leq
f^{G}/f^{\ast}$ which is shown to be $L=1/2$ in \cite{Fisher1978}. When the
objective function $f$ is \emph{monotone submodular} (which applies to
coverage metrics), then it has been shown that $L=(1-\frac{1}{e})$
\cite{Nemhauser1978} and becomes $L=(1-(1-\frac{1}{N})^{N})$ when the
allowable maximum number of agents is constrained to $N$. Recent work
\cite{Conforti1984},\cite{Wang2016, Liu2018} has further improved these
performance bounds by exploiting the specific nature of the monotonicity (also
known as \textit{curvature} properties) of the specific objective function. By
using these improved bounds, the solutions to a variety of optimal coverage
problems in \cite{Sun2019} have been shown to often approach $L=1$, i.e., to
yield almost globally optimally solutions.

Our contributions in this paper are threefold. First, we formulate the problem
of determining an optimal team composition under a heterogeneous set of agents
as a resource allocation problem without introducing additional non-convexity
features to it. In particular, instead of treating the (discrete) number of
agents in each class as a decision variable, we associate this number with the
(continuous) sensing capacity of the agents in each class; hence, an
allocation of zero sensing capacity implies a virtual (or non-existing) agent.
In our problem formulation, instead of imposing a hard cardinality constraint,
an $l_{1}$ norm penalty in the objective function is employed to induce
sparsity and prevent any new non-convexity from being introduced.

Secondly, for the coverage component of the objective function (i.e., without
the aforementioned penalty term), a greedy algorithm is used and two new
improved performance bounds are derived based on the concepts of partial
curvature \cite{Liu2018}, total curvature, and greedy curvature
\cite{Conforti1984}.

Finally, we propose a distributed projected gradient ascent algorithm to solve
the overall optimal team composition problem. The key to this algorithm is the
proper selection of an \emph{initial condition} which is characterized by a
provable lower bound. Thus, we first use a greedy method to generate a
candidate solution to the underlying coverage component of the problem which
always contains all the available agents. This is used as the initial
condition to solve the main problem (combining coverage and system cost). In
doing so, a distributed projected gradient ascent scheme is used whose final
solution recovers both the integer and real variables associated with the
problem which respectively define the optimal team composition and the optimal
agent locations.

%\textcolor{red}{
%As the continuation of our previous work \cite{Sun2019, Zhong2011}, the above contributions are significantly different. The composition of heterogeneous teams is first considered in this work. Note that such team composition problem is combinatorial and NP-hard. Moreover, the heterogeneity considered here brings challenges to both the greedy algorithm and the performance bounds, which are addressed there. Besides, two extra tighter bounds are proposed here compared to \cite{Sun2019} and a technique is proposed to compute a performance bound corresponding to the obtained final solution (instead of the initial one from the greedy algorithm). At last, while \cite{Zhong2011} also employed the distributed gradient ascent algorithm, the first two contributions are not taken into account there.
%}

Relative to our previous work \cite{Sun2019, Zhong2011}, here we  consider a significantly different problem and
make a number of key contributions to the coverage control problem with heterogeneous agents.
A crucial difference in this work compared to both \cite{Sun2019, Zhong2011} is that we do not assume that a
given number of agents is to be deployed; rather, we seek to determine the number of agents (subject to an upper bound constraint) and
optimal team composition (not only  the optimal agent locations), which is a combinatorial NP-hard problem.
Moreover, heterogeneity considered in this  work brings challenges to the aforementioned greedy algorithm, to the
associated performance bounds, and to the process of determining an optimal team composition - all of which are
addressed here. Finally, two new tighter performance bounds are derived compared to those in \cite{Sun2019}.

%In doing so, a distributed projected gradient ascent scheme is used, and its final solution defines the composition of the optimal multi-agent team.

%%Arrangement of the paper
%The rest of the paper is organized as follows. The optimization problem for
%determining the optimal team composition is presented in Section
%\ref{sec:problem}. For the coverage objective function component of the main
%problem, a greedy algorithm is presented in Section \ref{sec:submodularity}
%along with some new performance bound guarantees based on submodularity
%theory. Subsequently, a distributed projected gradient ascent algorithm is
%proposed in Section \ref{sec:gradient} to solve the overall optimal team
%composition problem. Numerical results are included in Section
%\ref{sec:numerical} to validate the effectiveness of the proposing algorithm
%and Section \ref{sec:conclusions} concludes the paper.

%Arrangement of the paper

The rest of the paper is organized as follows. The optimization problem for determining the optimal team
composition is formulated in Section \ref{sec:problem}. Then, to obtain a good initial condition to solve this
optimization problem, a greedy algorithm is presented in Section \ref{sec:submodularity}, along with some
performance bound guarantees. Subsequently, to completely solve the formulated optimization problem, a distributed
projected gradient ascent process is proposed in Section \ref{sec:gradient}, along with some theoretical results
regarding the nature of its terminal solution. Numerical results are included in Section \ref{sec:numerical} to validate
the effectiveness of the proposed solution technique. Finally, Section \ref{sec:conclusions} concludes the paper.

%\vspace{-0.25cm}

\subsection{Preliminaries}
Some notations used throughout this paper are introduced here. The $n$-dimensional Euclidean space is denoted by $\mathbb{R}^n$. Lowercase letters are used to denote vectors (E.g. $x\in\mathbb{R}^n$) and bold (and lowercase) letters are used to denote matrices (E.g. $\mathbf{s}\in \mathbb{R}^{N \times 2}$) while uppercase letters are used to denote set variables. Moreover, $\vert \cdot\vert$ and $\Vert \cdot \Vert$ denote the cardinality of a set variable and the $l_2$ norm of a vector respectively.

\section{Problem Formulation}

\label{sec:problem}

\label{sec:problem copy(1)} We begin with a brief review of the the
multi-agent coverage problem (see
\cite{caicedo2008performing,Zhong2011,cortes2004coverage}). The mission space
$\Omega\subseteq\mbox{$\mathbb R$}^{2}$ is modeled as a convex compact
polygon. For non-convex polygons $\Omega_{1}$, such as the self-intersecting
ones, we make $\Omega$ the convex hull of $\Omega_{1}$, while $\Omega
\setminus\Omega_{1}$ defines obstacles that agents have to avoid. Let
$R(x):\mbox{$\mathbb R$}^{2}\rightarrow\mathbb{R}$ be an event density
function such that $R(x)\geq0,\forall x\in\Omega$ and $\int_{\Omega
}R(x)dx<\infty$ such that $R(x)$ represents the relative importance of a point
$x\in\Omega$. Obstacles present in the mission space can both limit the
movement of agents and interfere with their sensing capacities. Such obstacles
are modeled as non-intersecting polygons $M_{1},\ldots,M_{m}$ and their
interiors are forbidden regions for the agents. As a result, the feasible
(safety) region is $F=\Omega\setminus(\mathring{M}_{1}\cup\ldots\cup
\mathring{M}_{m})$, where $\mathring{M}$ is the interior of $M$.

\begin{figure}[b]
\centering
\includegraphics[width = 2.7in]{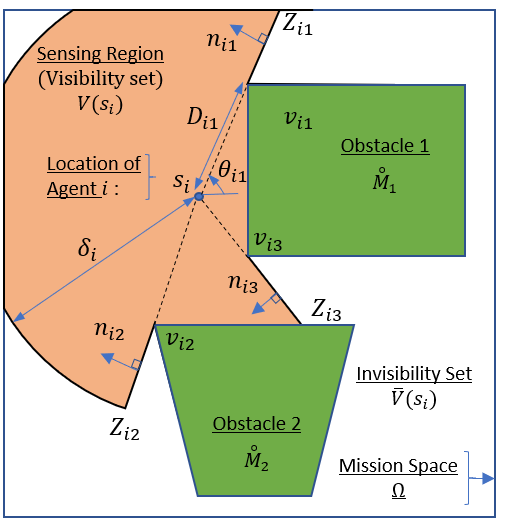} \caption{Mission
space with obstacles.}%
\label{f:illustration}%
\end{figure}

With $N$ as the maximum possible number of agents, we have $\mathbf{s}%
=[s_{1}^{T},\ldots,s_{N}^{T}]^{T}\in\mbox{$\mathbb R$}^{N \times2}$ denoting
the locations of the $N$ agents with each $s_{i}\in\mbox{$\mathbb R$}^{2}%
,\forall i=1,\ldots,N$. Then, the following sensing model is adopted. For any
point $x\in\Omega$ and a certain agent at $s_{i}$, there are two issues
affecting if the agent can detect an event occurring at $x$. First, the agent
is characterized by a sensing region defined as $\Omega_{i}=\{x|\Vert
x-s_{i}\Vert\leq\delta_{i}\}$, where $\delta_{i}$ is the sensing range.
Secondly, obstacles prevent a signal at $x$ from reaching $s_{i}$. This is
described by the condition $\eta s_{i} + (1-\eta)x\in F$, $\eta\in\lbrack
0,1]$, i.e., the segment connecting $x$ and $s_{i}$ must be contained in the
feasible region. Then, the visibility set of $s_{i}$ is defined as
$V(s_{i})=\Omega_{i}\cap\{x|\eta s_{i}+(1-\eta)x\in F\}$ and the invisibility
set $\bar{V}(s_{i})$ is the complement of $V(s_{i})$ in $F$, i.e., $\bar
{V}(s_{i})=F\setminus V(s_{i})$. An illustration of $V(s_{i})$ is shown in
Fig. \ref{f:illustration}.

The probability that agent $i$ detects an event at $x$ in an unconstrained
environment is given by
\begin{equation}
{p_{i}}(x,s_{i})=p_{i0}e^{-\lambda_{i}\Vert x-s_{i}\Vert}
\label{Def_detectionprob}%
\end{equation}
where $p_{i0}\in(0,1]$ is the agent's sensing capacity and $\lambda_{i}>0$ is
a sensing decay (attenuation) factor. As discussed in the introduction,
different ${p_{i}}(x,s_{i})$ specified by $p_{i0}$, $\delta_{i}$ and
$\lambda_{i}$ will lead to a heterogeneous multi-agent system. In a mission
space with constraints, the agent's detection probability becomes:%
\begin{equation}
\hat{p_{i}}(x,s_{i})=%
\begin{cases}
{p_{i}}(x,s_{i}) & \text{ if }x\in{V}(s_{i}),\\
0 & \text{ otherwise.}%
\end{cases}
\label{eq:detection}%
\end{equation}
Finally, assuming detection independence among the $N$ agents, the joint
detection probability of an event at $x$ is given by
\[
\hat{P}(x,\mathbf{s})=1-\Pi_{i=1}^{N}(1-\hat{p_{i}}(x,s_{i})).
\]
As formulated in \cite{Zhong2011}, the optimal multi-agent coverage problem
is
\begin{align}
\max_{\mathbf{s}}\ \  &  H(\mathbf{s})=\int_{\Omega}R(x)\hat{P}(x,\mathbf{s}%
)dx\label{eq:coverage1}\\
\text{s.t.}\ \  &  s_{i}\in F,\text{ \ \ }i=1,\ldots,N,\nonumber
\end{align}
where the number of the agents $N$ is a predetermined constant. When $N$ is in
fact an additional decision variable constrained by the cost of agents, we
proceed by capturing the trade-off between improved performance, which
monotonically increases with $N$, and agent cost as follows. Letting $N$ be
the the \emph{maximum possible number} of agents to consider, we formulate a
resource (sensing capacity) allocation problem:
\begin{align}
\max_{\mathbf{s},t}\ \  &  H(\mathbf{s},t)=\int_{\Omega}R(x)P(x,\mathbf{s}%
,t)dx-\beta\sum_{i=1}^{N}t_{i}\label{eq:coverage2}\\
\text{s.t.}\ \  &  s_{i}\in F,\ \ t_{i}\in\{0,1\},\ \ i=1,\ldots,N,\nonumber
\end{align}
with,
\begin{equation}
P(x,\mathbf{s},t)=1-\Pi_{i=1}^{N}(1-{t_{i}}\hat{p_{i}}(x,s_{i})).
\label{eq:joint2}%
\end{equation}
In (\ref{eq:coverage2}), $t=[t_{1},t_{2},\ldots,t_{N}]^{T}$ and $t_{i}$ is a
binary decision variable associated with agent $i$. The term $\beta\sum
_{i=1}^{N}t_{i}$ denotes the cost of deploying $N$ agents, where $\beta\geq0$
is a weight capturing the cost of each agent (assumed to be the same in this
formulation). In order to ensure a properly normalized objective function,
$\beta$ must be selected to be consistent with the following convex
combination of objectives:
\[
\tilde{H}(\mathbf{s},t)=w_{1}\frac{1}{\int_{\Omega}R(x)dx}\int_{\Omega
}R(x)P(x,\mathbf{s})dx-(1-w_{1})\frac{1}{N}\sum_{i=1}^{N}t_{i},
\]
where $w_{1}\in(0,1]$ (resp. $1-w_{1}$) and $\int_{\Omega}R(x)dx$ (resp. $N$)
are weights associated with the coverage performance metric (resp. cost
function). Observing that each component above is properly normalized in
$[0,1]$, we can adopt (\ref{eq:coverage2}) as long as $\beta$ is selected so
that
\begin{equation}
\beta=\frac{1-w_{1}}{w_{1}}\frac{\int_{\Omega}R(x)dx}{N}. \label{eq:beta_def}%
\end{equation}

Note that with $t_{i}\in\{0,1\}$, the agent heterogeneity in $\hat{p_{i}}$
(which depends on the values of $p_{i0}$ and $\lambda_{i}$ in
(\ref{Def_detectionprob}) and on the sensing range $\delta_{i}$) is not
included in the formulation (\ref{eq:coverage2}). In order to capture this
aspect of the problem, we relax the binary nature of $t_{i}$ by allowing it to
be a continuous variable $t_{i}\in\lbrack0,1]$. We then rewrite the detection
probability in (\ref{Def_detectionprob}) as $t_{i}p_{i0}e^{-\lambda_{i}\Vert
x-s_{i}\Vert}$ so that $t_{i}$ acts as a discount factor for the sensing
capacity $p_{i0}$. Accordingly, (\ref{eq:detection}) is modified to
\begin{equation}
\hspace{-0.4cm}\bar{p_{i}}(x,s_{i},t_{i})=%
\begin{cases}
t_{i}p_{i0}e^{-\lambda_{i}\Vert x-s_{i}\Vert} & \text{ if }x\in{V}(s_{i}),\\
0 & \text{ otherwise.}%
\end{cases}
\label{eq:detectionwithti}%
\end{equation}
and the definition of the joint detection probability in \eqref{eq:joint2}
becomes
\[
\bar{P}(x,\mathbf{s},t)=1-\Pi_{i=1}^{N}(1-\bar{p_{i}}(x,s_{i},t_{i})).
\]
With $\bar{P}(x,\mathbf{s},t)$ as defined above, we now extend
\eqref{eq:coverage2} to
\begin{align}
\max_{\mathbf{s},t}  &  \int_{\Omega}R(x)\bar{P}(x,\mathbf{s},t)dx-\beta
\sum_{i=1}^{N}t_{i}\label{eq:coverage3}\\
\text{s.t.}\ \  &  s_{i}\in F,\ \ t_{i}\in\lbrack0,1],\ \ i=1,\ldots
,N.\nonumber
\end{align}
%where $|\bullet|_0$ denotes the cardinality, i.e., number of non-zero elements, of a vector.
However, this formulation still does not capture the fact that agents with
different sensing parameter values $p_{i0}$, $\delta_{i}$ and $\lambda_{i}$
have different costs. Therefore, let $\gamma_{i}(p_{i0},\lambda_{i},\delta
_{i})$ denote the cost of agent $i$ and let us still keep $\beta$ as a weight
indicating the overall relative importance of cost relative to the coverage
performance expressed by the first term in the objective function. Omitting
the dependence of $\gamma_{i}$ on the sensing parameters, we now formulate the
problem:%
\begin{align}
\max_{\mathbf{s},t}\ \  &  H(\mathbf{s},t)=\int_{\Omega}R(x)\bar
{P}(x,\mathbf{s},t)dx-\beta\sum_{i=1}^{N}\gamma_{i}t_{i}\label{eq:coverage5}\\
\text{s.t.}\ \  &  s_{i}\in F,\ \ t_{i}\in\lbrack0,1],\ \ i=1,\ldots
,N.\nonumber
\end{align}
Clearly, heterogeneity here is captured in two ways: first, by imposing a
different cost $\gamma_{i}$ to each agent and second by associating a
different sensing capacity $t_{i}p_{i0}$ in (\ref{eq:detectionwithti}) to each
agent, assuming that such capacity is adjustable. More importantly, while the
binary constraint in \eqref{eq:coverage2} is removed, the $l_{1}$ norm used in
(\ref{eq:coverage5}) is a regularization term which is well known to induce
sparsity (e.g., in the use of machine learning algorithms such as
LASSO~\cite{tibshiranit1996regression}). The implication is that solutions of
this problem will tend to include values $t_{i}=0$ for several agents in
seeking cost-effective team compositions. This is both theoretically proven in
Theorem \ref{Th:optimalt_ivalues}, Section \ref{sec:gradient} and
experimentally validated using numerical results in Section
\ref{sec:numerical}.
%(this is clearly seen in our numerical results in Section \ref{sec:numerical}).

As in the case of (\ref{eq:coverage2}), the objective function in
(\ref{eq:coverage5}) needs to be properly normalized. To accomplish this while
also providing a physical interpretation to the cost coefficients $\gamma_{i}%
$, recall that $\Omega_{i}=\{x|\Vert x-s_{i}\Vert\leq\delta_{i}\}$ represents
the sensing region of agent $i$ and define the \textit{sensing capability} of
this agent as
\[
\kappa_{i}=\int_{\Omega_{i}}\hat{p}_{i}(x,s_{i})dx,
\]
where $s_{i}\in\mbox{$\mathbb R$}^{2}$ can be any point in the boundless and
obstacle-free space, hence $\kappa_{i}$ is independent of $s_{i}$; it depends
only on the sensing parameters $p_{i0}$, $\lambda_{i}$, and $\delta_{i}$. In
fact, for the exponential sensing function given in \eqref{Def_detectionprob},
a closed-form expression for $\kappa_{i}$ can be obtained as
\[
\kappa_{i}=\frac{2\pi p_{i0}}{\lambda_{i}^{2}}[1-(1+\lambda_{i}\delta
_{i})e^{-\lambda_{i}\delta_{i}}].
\]
Now, assuming the cost $\gamma_{i}$ associated with agent $i$ is proportional
to its sensing capability $\kappa_{i}$, we write:
\begin{equation}
\gamma_{i}=w_{2i}\kappa_{i}, \label{Eq:gamma_i_def}%
\end{equation}
where $w_{2i}\in(0,1]$ is a prespecified \textit{agent cost weight}. Finally,
we update the definition of the normalization factor $\beta$ in
\eqref{eq:beta_def} as follows:
\begin{equation}
\beta=\frac{1-w_{1}}{w_{1}}\frac{\int_{\Omega}R(x)dx}{\sum_{i=1}^{N}\gamma
_{i}}. \label{eq:beta_def_2}%
\end{equation}

With that, we can compute all the parameters/coefficients behind the formulated optimal agent team composition problem \eqref{eq:coverage5}, when the agent sensing capabilities and weights (i.e., $w_1$ and $w_{2i},\ \forall i$) are given. Therefore, the problem formulation is now complete.

%\vspace{-0.25cm}

\section{Greedy Algorithm and Submodularity Theory for Coverage Problems}

\label{sec:submodularity}

In order to obtain an initial solution to the problem in \eqref{eq:coverage5},
we first consider the problem given in \eqref{eq:coverage1} where the
objective is limited to maximizing the coverage using all the available
agents. Let us start by adopting the generic greedy method proposed in
\cite{Sun2019} and seek to improve upon the performance bounds provided in
\cite{Sun2019} by exploiting the curvature concepts proposed in
\cite{Conforti1984,Liu2018}.

\subsection{Set-function approach to the basic coverage problem}

In order to take advantage of the submodular structure of $H(\mathbf{s})$ in
(\ref{eq:coverage1}), we first uniformly discretize the continuous feasible
space $F$ to form a \textit{ground-set} $F^{D}=\{x_{1},x_{2},\ldots,x_{n}\}$
with each $x_{i}\in F$. These $x_{i}$ values can be thought of as feasible
points where an agent can be placed. Note that the cardinality $|F^{D}|$ of
the ground-set is $|F^{D}|=n$. As the next step, a \textit{set-variable} is
defined as $S=\{s_{1},s_{2},\ldots\}$ to represent the initial placement for
each agent. Typical constraints on selecting $S$ include the fact that each
$s_{i}$ should be chosen from the ground-set $F^{D}$ and the total number of
agents should be constrained to $N$. Therefore, the set-constraint
$S\in\mathcal{I}$, where $\mathcal{I}=\{A:A\subseteq F^{D},|A|\leq N\}$ is
used. Typically, a set-constraint of this form is called a \textit{uniform
matroid constraint of rank }$\mathit{N}$ where the pair $\mathcal{M}%
=(F^{D},\mathcal{I})$ is known as a \textit{uniform matroid}.

Furthermore, throughout this section, we approximate the coverage objective
function $H(\mathbf{s})$ in \eqref{eq:coverage1} by a set-function $H(S)$,
where $H:\mathcal{I}\rightarrow\mathbb{R}$ and
\begin{equation}
H(S)=\int_{\Omega}R(x)(1-\prod_{s_{i}\in S}\left[  1-\hat{p}_{i}%
(x,s_{i})\right]  )dx. \label{eq:setCoverageObjective}%
\end{equation}
Therefore, $H(S)$ now represents the coverage objective value achieved by the
agent placement defined by the set-variable $S$. In this new framework, a
set-function version of the original coverage problem in \eqref{eq:coverage1}
can be written as
\begin{equation}
\max_{S}\ \ H(S)\text{ \ \ s.t.}\ \ S\in\mathcal{I}.
\label{eq:setMaximization}%
\end{equation}

\subsection{Greedy algorithm}

Due to the combinatorial search space size, an exact solution to
\eqref{eq:setMaximization} is challenging to obtain. However, a candidate
solution can be obtained using a simple greedy algorithm and is referred to as
a \textit{greedy solution}. Here, we follow the greedy method given in
Algorithm \ref{Alg:Greedy} to obtain the corresponding greedy solution.

The marginal gain in the coverage objective due to adding a new agent at point
$x_{i}\in F^{D}$ to an existing agent set $A$ is denoted by $\Delta
H(x_{i}|A)$ where
\begin{equation}
\Delta H(x_{i}|A)=H(A\cup\{x_{i}\})-H(A).
\end{equation}
This is also known as the \textit{discrete derivative} of the set function
$H(S)$ at $S=A$ in the direction $x_{i}$ \cite{Conforti1984}. It can be shown
through a straightforward evaluation based on (\ref{eq:setCoverageObjective})
that
\begin{equation}
\Delta H(x_{i}|A)=\int_{F}R(x)p_{i}(x,s_{i})\prod_{s_{j}\in A}\left[
1-\hat{p}_{j}(x,s_{j})\right]  dx. \label{eq:discreteDerivative}%
\end{equation}

\begin{algorithm}[!h]
\caption{Greedy Method for Solving \eqref{eq:setMaximization}}
\label{Alg:Greedy}
\begin{algorithmic}[1]
\State \textbf{Inputs: }{$N, F^D$ (Recall $\mathcal{I}:= \{A:A \subseteq F^D,\ \vert A \vert \leq N \}$).}
\State \textbf{Outputs: } {Greedy solution $S^G$.}
\State $S := \emptyset$; $i := 1$;
\While{$i \leq N$}
\State $s_i := \arg \max_{\{x_i: (S \cup \{x_i\}) \in \mathcal{I} \}}
\left(\Delta H(x_i \vert S)\right)$;
\State $S := S \cup \{s_i\}$;
\EndWhile
\State $S^G := S$; \textbf{Return};
\end{algorithmic}
\end{algorithm}

Using the properties of the problem \eqref{eq:setMaximization}, we can now
derive several bounds allowing us to quantify how close the greedy solution is
to the globally optimal solution.

\subsection{Performance Bounds}

Consider the greedy solution of \eqref{eq:setMaximization} given by Algorithm
\ref{Alg:Greedy} as $S=S^{G}$. The \textit{performance ratio} of this greedy
solution is defined as $H(S^{G})/H(S^{\ast})$ where $S^{\ast}$ is the globally
optimal solution (of \eqref{eq:setMaximization}) and is generally unknown. A
\textit{performance bound} $L$ is defined as a theoretically imposed lower
bound to the performance ratio. Therefore,
\begin{equation}
\label{Eq:PerformanceBoundDef}L \leq\frac{H(S^{G})}{H(S^{\ast})} \leq1.
\end{equation}

It was proven in \cite{Sun2019} that the set-function $H(S)$ has two important
properties: \textit{submodularity} and \textit{monotonicity}. Therefore,
following the seminal paper \cite{Nemhauser1978}, the greedy solution to the
coverage problem in \eqref{eq:setMaximization} is characterized by the
performance bound $L=L_{C}$, where
\begin{equation}
L_{C}=(1-(1-\frac{1}{N})^{N}). \label{Eq:ConventionalBound}%
\end{equation}
We refer to \eqref{Eq:ConventionalBound} as the \emph{conventional}
performance bound.

\subsection{Curvature information}

For the class of coverage problems we are considering, it is shown in
\cite{Sun2019} that tighter performance bounds (i.e., performance bounds which
are closer to 1 than $L_{C}$) can be obtained using the \textit{curvature}
information of the objective function $H(S)$. Typically, any measure of
\textit{curvature} of a set function $f(A)$ provides additional information
about the nature of its growth when new elements are added to the set-variable
$A$. In other words, \textit{curvature} information characterizes the nature
of the monotonicity of $f(A)$. For example, the marginal gain of a coverage
objective set-function $H(S)$ (represented by $\Delta H(\cdot|S)$), can
drastically drop when elements are added to the set $S$. Due to this reason,
characterizing the set function's monotonicity (using curvature information)
can yield vital information about the effectiveness of greedy methods.

\subsubsection{Total Curvature}

The concept of \textit{total curvature} for generic submodular monotone
set-functions was introduced in \cite{Conforti1984}. When this concept is
applied to the class of coverage control problems, the total curvature of
$H(S)$ denoted by $\alpha_{T}$ is given by
\begin{equation}
\alpha_{T}=\max_{x_{i}:x_{i}\in F^{D}}\left[  1-\frac{\Delta H(x_{i}%
|F^{D}\backslash x_{i})}{\Delta H(x_{i}|\emptyset)}\right]  ,
\label{Eq:TotalCurvature}%
\end{equation}
where we use $\emptyset$ to denote the empty set. Further, \textquotedblleft%
\ $\cdot\backslash\cdot$\ " is used to denote the set-subtraction operation
(i.e., $A\backslash B=A\cap B^{c}$). The use of the prefix \textquotedblleft
total\textquotedblright\ comes from the fact that $\alpha_{T}$ is evaluated
based on the marginal gain $\Delta H(\cdot|S)$ when $S=\emptyset$ and when
$S=F^{D}\backslash s_{j}$ (i.e., at extreme ends of possible sets $S$).
Therefore, the total curvature measure tries to characterize the monotonicity
of $H(S)$ using its marginal gain evaluated at two extreme ends of choices for
$S$. In the context of real-valued functions defined on a finite interval, the
use of total curvature (for monotone submodular set functions) is analogous to
attempting to characterize the shape of a monotonically increasing curve with
monotonically decreasing gradient, using only its gradient at its two endpoints.

Using \eqref{eq:discreteDerivative}, \eqref{eq:setCoverageObjective}, and the
knowledge of $F^{D}$, the total curvature $\alpha_{T}$ of the set-function
$H(S)$ can be explicitly evaluated. In \cite{Conforti1984}, it is shown that
when maximizing a submodular monotone set function with a total curvature
$\alpha_{T}$, the greedy solution will follow the performance bound $L =
L_{T}$ where
\begin{equation}
\label{Eq:TotalCurvatureBound}L_{T} = \frac{1}{\alpha_{T}}\left[  1-\left(
\frac{N-\alpha_{T}}{N}\right)  ^{N}\right]  .
\end{equation}
This total curvature measure has been used in \cite{Sun2019} to establish
better performance bounds compared to the conventional bound $L_{C}$ in the
context of the coverage control problem in \eqref{eq:setMaximization}. Next,
we propose another curvature concept to obtain even tighter performance bounds
than $L_{T}$.

\subsubsection{Partial Curvature}

In \cite{Liu2018}, the concept of \textit{partial curvature} is proposed for
submodular monotone set functions which are defined under uniform matroid
constraints. Adopting this new concept, the partial curvature measure
associated with the coverage objective set-function $H(S)$ can be expressed as
$\alpha_{P}$ where
\begin{equation}
\alpha_{P}=\max_{(A,x_{i}):x_{i}\in A\in\mathcal{I}}\left[  1-\frac{\Delta
H(x_{i}|A\backslash x_{i})}{\Delta H(x_{i}|\emptyset)}\right]  .
\label{Eq:PartialCurvature}%
\end{equation}
As discussed in \cite{Liu2018}, the partial curvature delivers a better
characterization of the monotonicity of any generic set-function compared to
the total curvature. This improvement is due to the fact that only the
information obtained from the domain of the considered set-function is used -
which can be considerably smaller due to the uniform matroid constraint. The
importance of the partial curvature concept in the context of our coverage
problem can be explained as follows. For coverage problems, evaluating
$H(F^{D})$ so as to compute the total curvature in \eqref{Eq:TotalCurvature}
and then to impose the performance bound $L_{T}$ is problematic because the
domain of $H(\cdot)$ in the original optimization problem
\eqref{eq:setMaximization} is actually limited to size $N$ sets (i.e., by the
constraint $S\in\mathcal{I}$). This issue is critical when we consider
heterogeneous agents (in terms of sensing capabilities) and a finite set of
agents at our disposal to achieve the maximum coverage. In such situations,
$H(F^{D})$ is ill-defined and, therefore, the total curvature and the
respective performance bound $L_{T}$ cannot be evaluated. However, the
definition of the partial curvature in\eqref{Eq:PartialCurvature} will still
hold as it only requires evaluations of $H(\cdot)$ over the same domain (i.e.,
$S\in\mathcal{I}$).

\begin{rem}
When all the agents available are homogeneous (as opposed to the heterogeneous
situation discussed above) the definition of the coverage objective function
$H(\cdot)$ in \eqref{eq:setCoverageObjective} is flexible enough so that its
domain can be extended to $2^{F^{D}}$ (from $\mathcal{I}$). Thus, it enables
the evaluation of the total curvature measure in \eqref{Eq:TotalCurvature} and
the associated performance bound $L_{T}$. However, the effectiveness of the
bound $L_{T}$ is questionable since this has been computed using a larger
objective function domain ($2^{F^{D}}$) while the original optimization
problem in \eqref{eq:setMaximization} is considered over a smaller domain
$\mathcal{I}$. Therefore, it is natural to presume that the total
curvature-based performance bound $L_{T}$ can be further improved when the
optimization problem is over a smaller domain.
\end{rem}

Using \eqref{eq:discreteDerivative}, \eqref{eq:setCoverageObjective} and the
knowledge of $\mathcal{I}$, the partial curvature $\alpha_{P}$ in
\eqref{Eq:PartialCurvature} can be computed for the coverage problem. The
corresponding performance bound is denoted by $L = L_{P}$, where
\begin{equation}
\label{Eq:PartialCurvatureBound}L_{P} = \frac{1}{\alpha_{P}}\left[  1-\left(
\frac{N-\alpha_{P}}{N}\right)  ^{N}\right]  .
\end{equation}

\subsubsection{Greedy Curvature}

We also introduce the use of another curvature concept, the \textit{greedy
curvature}, which is proposed in \cite{Conforti1984} as an on-line method of
estimating a performance bound. The resulting performance bound depends on the
greedy solution $S^{G}$ itself. Note that the performance bounds discussed
thus far are not dependent on the obtained greedy solution but only the
objective function parameters (such as $\lambda_{i},\delta_{i}$ for all $i$)
and $N$, as well as the feasible space $F^{D}$.

If the greedy algorithm given in Algorithm \ref{Alg:Greedy} produces the
solution sets $\emptyset=S^{0}\subseteq S^{1}\subseteq S^{2}\subseteq
\cdots\subseteq S^{N}$ during the course of execution (where $S^{N}=S^{G}$),
then, the greedy curvature metric $\alpha_{G}$ is given by
\begin{equation}
\alpha_{G}=\max_{0\leq i\leq N-1}\left[  \max_{x_{j}\in F^{i}}\left(
1-\frac{\Delta H(x_{j}|S^{i})}{\Delta H(x_{j}|\emptyset)}\right)  \right]  ,
\label{Eq:GreedyCurvature}%
\end{equation}
where $F^{i}=\{x_{j}:x_{j}\in F^{D}\backslash S^{i},(S^{i}\cup\{x_{j}%
\})\in\mathcal{I}\}$ is the set of valid points considered for the placement
of the $(i+1)$\textsuperscript{th} agent during the $(i+1)$%
\textsuperscript{th} greedy iteration. Therefore, $\alpha_{G}$ can be computed
in parallel with the greedy method (without performing any additional
computations) unlike the previously discussed two cases. The corresponding
performance bound denoted by $L = L_{G}$ is
\begin{equation}
L_{G}= 1-\alpha_{G}(1-\frac{1}{N}). \label{Eq:GreedyCurvatureBound}%
\end{equation}
The main idea behind the greedy curvature concept is that the solution sets
generated during the greedy algorithm itself can be used to characterize the
monotonicity of the considered set-function and then to establish a
performance bound based on that information. Therefore, similar to the
observation made earlier regarding the feasibility of using the total
curvature-based performance bound $L_{T}$ for a heterogeneous set of agents,
the definition of the greedy curvature measure in \eqref{Eq:GreedyCurvature}
and the performance bound $L_{G}$ in \eqref{Eq:GreedyCurvatureBound} will
still hold in such cases.

\subsection{The Overall Performance Bound $L$}

Taking all the aforementioned performance bounds $L_{C},L_{T},L_{P}$ and
$L_{G}$ defined respectively in \eqref{Eq:ConventionalBound},
\eqref{Eq:TotalCurvatureBound}, \eqref{Eq:PartialCurvatureBound}, and
\eqref{Eq:GreedyCurvatureBound}, into account, an \textit{overall performance
bound} $L$ satisfying \eqref{Eq:PerformanceBoundDef} can be established as
\begin{equation}
L=\max{\{L_{C},L_{T},L_{P},L_{G}\}}. \label{Eq:OverallPerformanceBound}%
\end{equation}
Generally, $L_{C}\leq L_{T}\leq L_{P}$ \cite{Liu2018}. Also, recall that when
heterogeneous agents are involved, $L_{T}$ and $L_{G}$ are undefined.

\subsection{Numerical results for greedy method}

\begin{figure}[h]
\centering
\begin{subfigure}{0.23\columnwidth}
\includegraphics[width=\textwidth]{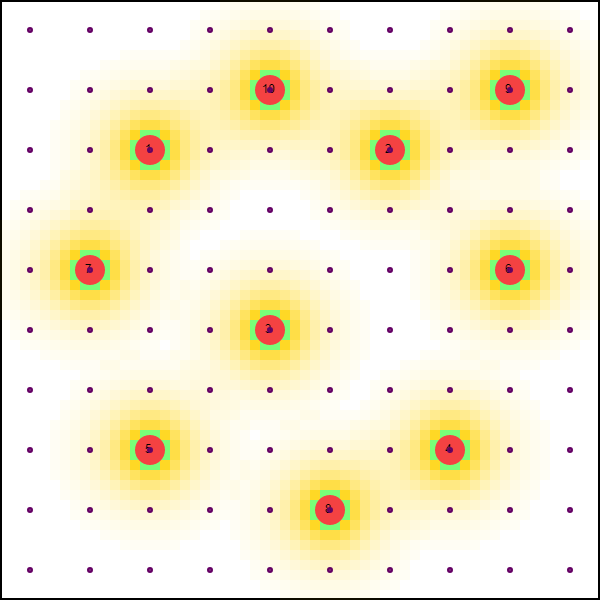}
\caption{Blank,\\$\delta_i = 100,\\ \lambda_i = 0.04$}
\label{fig:Bsetting}
\end{subfigure}
\begin{subfigure}{0.23\columnwidth}
\includegraphics[width=\textwidth]{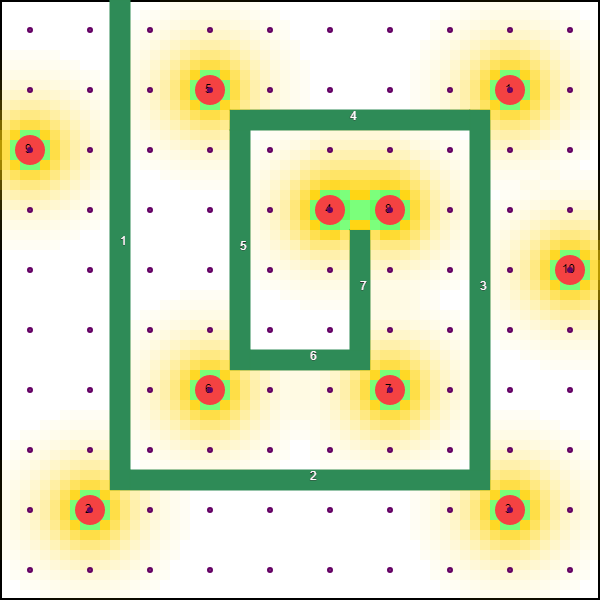}
\caption{Maze\\$\delta_i = 100,\\ \lambda_i = 0.04$}
\label{fig:Msetting}
\end{subfigure}
\begin{subfigure}{0.23\columnwidth}
\includegraphics[width=\textwidth]{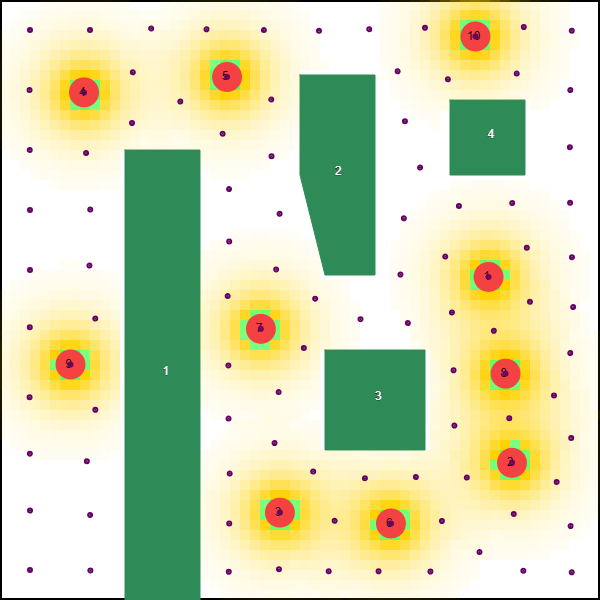}
\caption{Global1\\$\delta_i = 100,\\ \lambda_i = 0.04$}
\label{fig:G1setting}
\end{subfigure}
\begin{subfigure}{0.23\columnwidth}
\includegraphics[width=\textwidth]{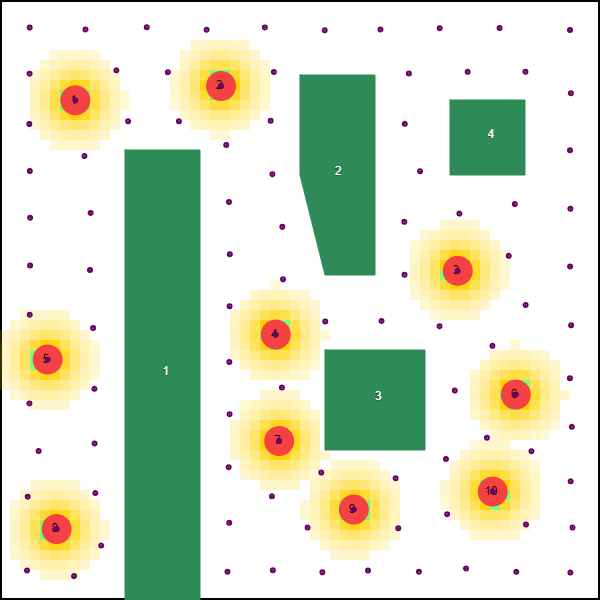}
\caption{Global2\\$\delta_i = 50,\\ \lambda_i = 0.05$}
\label{fig:G2setting}
\end{subfigure}
\caption{Different problem settings and their greedy solutions for $N=10$. Red
dots are greedy agent locations, black dots represent the ground set. Darker
colored areas have greater coverage, and green colored shapes are obstacles.}%
\label{Fig:ProblemSettings}%
\end{figure}

We now investigate the behavior of the proposed partial curvature and greedy
curvature-based performance bounds $L_{P}, L_{G}$ compared to the
conventional and total curvature performance bounds $L_{C}$, $L_{T}$. Four
different representative problem settings were considered as shown in Fig.
\ref{Fig:ProblemSettings}. Under each of these settings, the aforementioned
performance bounds were evaluated for different values of the total allowable
number of agents $N$.

%\begin{figure}[h]
%\centering
%\begin{subfigure}{0.45\columnwidth}
%\includegraphics[width=\textwidth]{figures/Bplot}
%\caption{}
%\label{fig:Bplot}
%\end{subfigure}
%\begin{subfigure}{0.45\columnwidth}
%\includegraphics[width=\textwidth]{figures/Mplot}
%\caption{}
%\label{fig:Mplot}
%\end{subfigure}
%\begin{subfigure}{0.45\columnwidth}
%\includegraphics[width=\textwidth]{figures/G1plot}
%\caption{}
%\label{fig:G1plot}
%\end{subfigure}
%\begin{subfigure}{0.45\columnwidth}
%\includegraphics[width=\textwidth]{figures/G2plot}
%\caption{}
%\label{fig:G2plot}
%\end{subfigure}
%\caption{Performance bounds (as a function of $N$): (\romannum{1})
%Conventional $L_{C}$, (\romannum{2}) Total curvature $L_{T}$, (\romannum{3})
%Partial curvature $L_{P}$, and (\romannum{4}) Greedy curvature $L_{G}$, for
%the four problem settings in Fig. \ref{Fig:ProblemSettings}.}%
%\label{Fig:PerformanceBounds}%
%\end{figure}

\begin{figure}[h]
\centering
\begin{subfigure}{0.45\columnwidth}
\includegraphics[width=\textwidth]{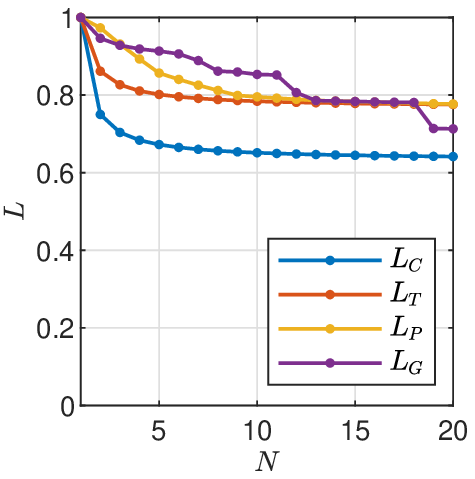}
\caption{}
\label{fig:Bplot}
\end{subfigure}
\begin{subfigure}{0.45\columnwidth}
\includegraphics[width=\textwidth]{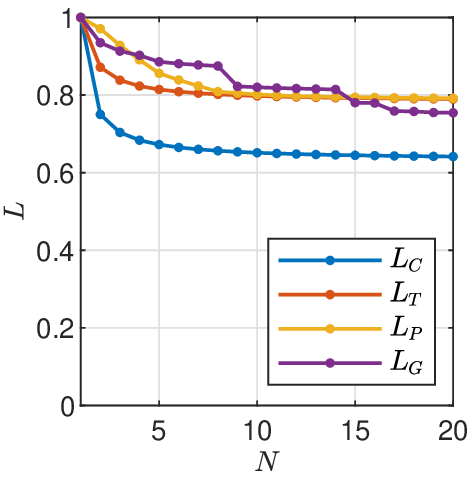}
\caption{}
\label{fig:Mplot}
\end{subfigure}
\begin{subfigure}{0.45\columnwidth}
\includegraphics[width=\textwidth]{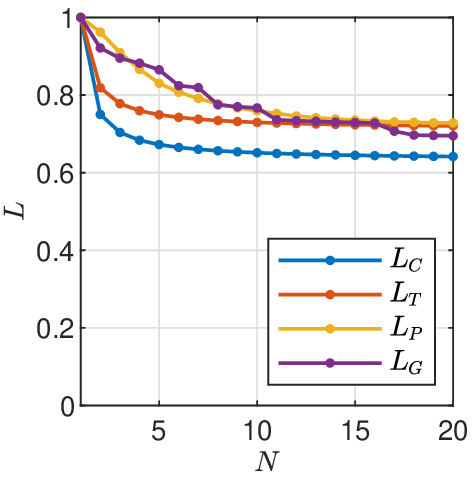}
\caption{}
\label{fig:G1plot}
\end{subfigure}
\begin{subfigure}{0.45\columnwidth}
\includegraphics[width=\textwidth]{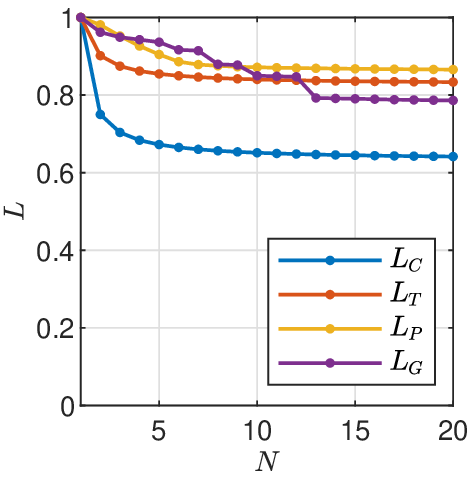}
\caption{}
\label{fig:G2plot}
\end{subfigure}
\caption{Performance bounds (as a function
of $N$): (\romannum{1}) Conventional $L_{C}$, (\romannum{2}) Total curvature
$L_{T}$, (\romannum{3}) Partial curvature $L_{P}$, and (\romannum{4}) Greedy
curvature $L_{G}$, for the four problem settings in Fig.
\ref{Fig:ProblemSettings}.}%
\label{Fig:PerformanceBounds}%
\end{figure}

% From the obtained results shown in Fig. \ref{Fig:PerformanceBounds}, it is
% evident that the proposed use of partial curvature always delivers better
% bounds than the total curvature approach \cite{Conforti1984}. The proposed
% greedy curvature-based performance bound outperforms the conventional method
% \cite{Nemhauser1978} when $N$ takes moderate values (i.e., $N$ is around
% $4-20$). Note that both $L_{P},L_{T}$ always outperform $L_{G}$. However,
% $L_{G}$ is useful for computation-limited settings, as it does not require any
% additional computations compared to evaluating performance bounds such as
% $L_{T}$ or $L_{P}$.

From the obtained results shown in Fig. \ref{Fig:PerformanceBounds}, it is
evident that the proposed use of partial curvature always delivers better
bounds than the total curvature approach \cite{Conforti1984}. Similarly, the proposed use of greedy curvature provides better bounds than the total curvature approach \cite{Conforti1984} when $N$ takes moderate values (i.e., $N$ is around $2-20$). Moreover, $L_{G}$ is useful for computation-limited settings, as it does not require any
additional computations compared to evaluating $L_{T}$ or $L_{P}$.

As pointed out earlier, the performance bound $L_{T}$ is ill-defined when
considering heterogeneous agents. To avoid this problem, the experiments
reported above were limited to a homogeneous set of agents. However, it should
be emphasized that the definitions of the proposed performance bounds $L_{P}$
and $L_{G}$ are robust to agent heterogeneity, the situation considered in
section \ref{sec:numerical}. Therefore, in such heterogeneous situations,
using $L_{P}$ and/or $L_{G}$ will be the only way to obtain an improved
performance bound compared to the conventional bound $L_{C}$. Note that in such situations, the greedy algorithm given will require an additional inner loop to determine the optimal type of the agent to be deployed at each main greedy iteration.

We conclude this section by reminding the reader that the greedy process detailed above is needed so as to
generate an initial condition to the main optimization problem in \eqref{eq:coverage5}.
We have also discussed different performance bound computation techniques which
can characterize the closeness of these initial conditions to the global optimum.

%\vspace{-0.25cm}

\section{A Gradient Based Algorithm for Heterogeneous Multi-Agent Coverage
Problem}

\label{sec:gradient}

The greedy algorithm (Algorithm \ref{Alg:Greedy}) is limited to discrete
environments and a fixed predetermined agent number. Its value in solving the
actual problem of interest in \eqref{eq:coverage5} is twofold: $(i)$ Provide a
reasonable initial condition for a gradient-based algorithm used to solve
\eqref{eq:coverage5} which can significantly overcome the local-optimality
limitation of such an algorithm, and $(ii)$ Provide a lower bound for the
ultimate coverage performance we obtain.

In this section, we propose a distributed gradient-based algorithm similar to
that in~\cite{Zhong2011} aimed at solving \eqref{eq:coverage5}. We first
derive the derivatives of the objective function $H(s,t)$ with regard to the
variables $(s,t)$ for the gradient ascent update. Setting $s_{i}%
=(s_{ix},s_{iy})$, we begin with $\frac{\partial H(s,t)}{\partial s_{ix}}$
whose derivation was given in \cite{Zhong2011}:
\begin{align}
&  \frac{\partial H(\mathbf{s},t)}{\partial s_{ix}}=\int_{V(s_{i})}%
R(x)\Phi_{i}(x)\frac{\partial\bar{p}_{i}(x,s_{i},t_{i})}{\partial s_{ix}%
}dx\label{eq:derivaties}\\
&  +\underset{j\in\Gamma_{i}}{\sum}sgn(n_{ijx})\frac{\sin(\theta_{ij})}%
{D_{ij}}\int_{0}^{Z_{ij}}R(\rho(r))\Phi_{i}({\rho(r)}) \bar{p}_{i}%
(\rho(r),s_{i},t_{i})rdr,\nonumber
\end{align}
where
\begin{align*}
\Phi_{i}(x)  &  = \underset{k \in B_{i}}{\Pi}[1-\bar{p}_{k}(x,s_{k},t_{k})],\\
\frac{\partial\bar{p}_{i}(x,s_{i},t_{i})}{\partial s_{ix}}  &  = -\lambda
_{i}\bar{p}_{i}(x,s_{i},t_{i})\left(  \frac{(s_{i}-x)_{x}}{\Vert s_{i}%
-x\Vert}\right)  ,\\
\rho(r)  &  = \rho_{ij}(r) = \left(  \frac{v_{ij}-s_{i}}{D_{ij}}\right)
r+v_{ij},\ \ \mbox{and,}\\
D_{ij}  &  = ||v_{ij}-s_{i}||.
\end{align*}

%\textbf{\st{XXX For self containment of the paper, let us include the definitions which are missing here and refer to Fig. 1 for explanations.}}

In \eqref{eq:derivaties}, $sgn(\cdot)$ represents the signum function and the
subscript $x$ is used to represent the $x$-component of a two dimensional
vector. The second term in \eqref{eq:derivaties} is due to the linear shaped
boundary segments of the sensing region $V(s_{i})$ formed due to the obstacle
vertices $v_{ij}\in V(s_{i})$. Such linear segments are lumped into a set
$\Gamma_{i}=\{\Gamma_{i1},\Gamma_{i2},\ldots\}$ where each linear segment
$\Gamma_{ij}$ can be characterized by four parameters: (\romannum{1}) end
point $Z_{ij}$, (\romannum{2}) angle $\theta_{ij}$, (\romannum{3}) obstacle
vertex $v_{ij}$, and, (\romannum{4}) unit normal direction $n_{ij}$.
Therefore, $\Gamma_{ij}$ can be thought of as a four-tuple $\Gamma
_{ij}=(Z_{ij},\theta_{ij},v_{ij},n_{ij})$. All these geometric parameters (for
a generic setting) are illustrated in Fig. \ref{Fig:ProblemSettings}. Note
that we assume: (\romannum{1}) obstacles are polygonal, and, (\romannum{2})
sensing power at the edge of the sensing region is negligible. More detailed
definitions and derivations are omitted for brevity, and interested readers
are referred to~\cite{Zhong2011}.

A similar expression can be obtained for $\frac{\partial H(s,t)}{\partial
s_{iy}}$. As~detailed in \cite{Zhong2011}, the agent locations are assumed not
to coincide with a reflex vertex, a polygonal inflection, or a bi-tangent
where $H(s,t)$ is not differentiable (if such points have to be taken into
consideration, then a subgradient can be used as an alternative to the gradient).

Additionally, the derivative $\frac{\partial H(s,t)}{\partial t_{i}}$ is
obtained as follows:
\begin{align}
\label{eq:derivaties_t}\frac{\partial H(\mathbf{s},t)}{\partial t_{i}} =
\underbrace{\int_{V(s_{i})}R(x)\Phi_{i}(x) p_{i}(x,s_{i})dx}%
_{{\mbox{Local Coverage}}} \ \ - \underbrace
{\vphantom{\left(\frac{dummy}{frac}\right) } \beta\gamma_{i}.}%
_{\mbox{Local Cost}}%
\end{align}
Here, the integration and differentiation are interchangeable since
$P(x,\mathbf{s})$ is a continuous differentiable function of $t_{i}$. The
first term in \eqref{eq:derivaties_t} represents a \textit{local coverage}
level achieved by the agent $i$ in its sensing region $V(s_{i})$. This local
coverage level depends on the state variables $(\mathbf{s},t)$ and is always
positive. The second term in \eqref{eq:derivaties_t} represents a
\textit{local cost} resulting from agent cost $\gamma_{i}$ and the
normalization factor $\beta$. Note that this local cost value is a predefined
positive constant for each agent. This multi-objective interpretation of
\eqref{eq:derivaties_t} can be used to conclude that when the aforementioned
local coverage level is less than the (fixed) local cost, the state variable
$t_{i}$ should be decreased to improve the global objective $H(\mathbf{s},t)$,
and vice versa.

%Here, the integration and differentiation are interchangeable since $P(x,\mathbf{s})$ is a continuous differentiable function of $t_{i}$. The first term in \eqref{eq:derivaties_t} denoted as $H_i$ represents a local coverage level achieved by the agent $i$ in its sensing region $V(s_i)$. This $H_i$ value depends on the state variables $(\mathbf{s},t)$ and $H_i>0 \ \forall (\mathbf{s},t)$. The second term in \eqref{eq:derivaties_t} represents the effect of agent cost $\gamma_i$ and it is a predefined constant for each agent. This multi-objective interpretation of \eqref{eq:derivaties_t} can be used to conclude that when $H_i \leq \beta \gamma_i$, (i.e., when the local coverage is less than the fixed local cost), $t_i$ should be decreased to improve the global objective $H(\mathbf{s},t)$, and vice versa.

Algorithm \ref{alg:PGA} is a Projected Gradient Ascent (PGA) algorithm for
solving \eqref{eq:coverage5} which utilizes the gradients derived in
\eqref{eq:derivaties} and \eqref{eq:derivaties_t}.
{As seen in Algorithm
\ref{alg:PGA}, a gradient ascent update is first implemented in
\eqref{eq:update}, where $\eta_s^{(k)}>0,\ \eta_t^{(k)}>0$ are the step sizes chosen based on standard technical conditions \cite{Bertsekas2016} (more application-specific details on the step size selection can be found in \cite{Welikala2019Ax}).}
Subsequently, the projection mechanisms are applied to guarantee the
satisfaction of all constraints. The projection $\Pi_{A}(x)$ of $x
\in\mbox{$\mathbb R$}^{n}$ onto a set $A \subseteq\mbox{$\mathbb R$}^{n}$ is
formally defined as
\begin{equation}
\label{Eq:Projection_Def}\Pi_{A}(x) \triangleq\arg\min_{y \in A} \Vert
x-y\Vert_{2}.
\end{equation}
For $s_{i}\in F$, if the update direction (i.e., $\frac{\partial
H(\mathbf{s},t)}{\partial s_{i}}$) is pointing directly into an obstacle's
boundaries, then the update direction is projected onto the boundary itself
and thus prevents violation of the obstacle constraint. As for the bound
constraint for $t_{i}$, a projection onto the convex set $[0,1]$ is simply a truncation.

\begin{algorithm}
\caption{Projected Gradient Ascent (PGA) Algorithm for solving the problem in  \eqref{eq:coverage5}.}
\label{alg:PGA}
\begin{algorithmic}[1]
\State \textbf{Inputs: }{$\Omega,\ F,\ N,$ and, tolerances $\epsilon_s, \epsilon_t > 0$.}
\State \textbf{Initialize: }{$\mathbf{s}^0 := [S^G]$ (From Alg. \ref{Alg:Greedy}), $t^0 := [1,1,\ldots,1] \in \mbox{$\mathbb R$}^{n}$.}
\State \textbf{Outputs:} {$\mathbf{s}^{PGA}$, $t^{PGA}$.}
\For{$k = 0,1,2, \ldots \ $}
\State \textbf{Compute: } At $(\mathbf{s},t) = (\mathbf{s}^{(k)}, t^{(k)}),\ \forall i \in \{1,2,\ldots,N\},$
\begin{eqnarray}
\begin{aligned}\label{Eq:ComputedGradients}
&\frac{\partial H(\mathbf{s},t)}{\partial s_{i}}
= \left[\frac{\partial H(\mathbf{s},t)}{\partial s_{ix}}, \frac{\partial H(\mathbf{s},t)}{\partial s_{iy}}\right]; \\
&\frac{\partial H(\mathbf{s},t)}{\partial t_{i}};
\end{aligned}
\end{eqnarray}\Comment{Using \eqref{eq:derivaties} and \eqref{eq:derivaties_t}.}
\State \textbf{Update: } $(\mathbf{s}^{(k)},t^{(k)}),$ by, $\ \forall i \in \{1,2,\ldots,N\},$
\begin{eqnarray}\label{eq:update}
\begin{aligned}
\hat{s}_i^{\,(k+1)} &=& s_i^{(k)}
+ \eta_s^{(k)} \frac{\partial H(\mathbf{s},t)}{\partial s_{i}}; \\
\hat{t}_i^{\,(k+1)} &=& t_i^{(k)}
+ \eta_t^{(k)} \frac{\partial H(\mathbf{s},t)}{\partial t_{i}};
\end{aligned}
\end{eqnarray}
\State \textbf{Projection: } to get $(\mathbf{s}^{(k+1)},t^{(k+1)})$, $\ \forall i \in \{1,2,\ldots,N\},$
\begin{eqnarray}\label{eq:projection}
    \begin{aligned}
    {s}_i^{(k+1)} &=& \Pi_F(\hat{s}_i^{\,(k+1)});\\
    {t}_i^{(k+1)} &=& \Pi_{[0,1]}(\hat{t}_i^{\,(k+1)});
    \end{aligned}
\end{eqnarray}
\If{$\{ \Vert \mathbf{s}^{(k+1)}-\mathbf{s}^{(k)}\Vert \leq \epsilon_s$ and $\Vert t^{(k+1)}-t^{(k)} \Vert \} \leq \epsilon_t\}$}
\State {$(\mathbf{s}^{PGA},t^{PGA}) := (\mathbf{s}^{(k+1)},t^{(k+1)});$ \textbf{Return;}}
\EndIf
\EndFor
\end{algorithmic}
\end{algorithm}

%\vspace{-0.5cm}

\paragraph*{\textbf{Coverage performance of the PGA solution} $\mathbf{s}%
^{PGA}$}

For the initialization of the PGA algorithm given in Algorithm \ref{alg:PGA},
we use the greedy solution $\mathbf{s}^{(0)}=[S^{G}]$ obtained from Algorithm
\ref{Alg:Greedy} using: (\romannum{1}) The pre-specified discretized feasible
space $F^{D}$, and, (\romannum{2}) The complete set of agents (all $N$ of
them). The overall performance bound obtained using
\eqref{Eq:OverallPerformanceBound} under this initial configuration is
$L_{1}\leq\frac{H(\{\mathbf{s}^{(0)}\})}{H(S^{\ast})}$. Therefore, $L_{1}$
does not convey any information about the coverage performance of the obtained
PGA solution. This issue is addressed as follows (using the notation $[\cdot]$
and $\{\cdot\}$ to represent a conversion from a set to an array and vice versa).

Once the PGA solution $(\mathbf{s}^{PGA},t^{PGA})$ is obtained using Algorithm
\ref{alg:PGA}, it yields information on: (\romannum{1}) optimal agent
locations (i.e., $\mathbf{s}^{PGA}$), and, (\romannum{2}) optimal agent team
composition (i.e., $t^{PGA}$). This allows us to update the discretized
feasible space $F^{D}$ into $F^{D2}$ by inserting the agent coordinates found
in $\mathbf{s}^{PGA}$ such that $F^{D2}\triangleq F^{D}\cup\{\mathbf{s}%
^{PGA}\}$. Next, we re-evaluate the greedy algorithm considering only the
agents in the optimal team and using the modified discretized feasible space
$F^{D2}$. Now, if the corresponding greedy solution is $S^{G2}$ and the
overall performance bound is $L_{2}$ (obtained from
\eqref{Eq:OverallPerformanceBound}), following \eqref{Eq:PerformanceBoundDef}
we can write $L_{2}\leq\frac{H(S^{G2})}{H(S^{\ast})}$. This relationship
together with $H(\{\mathbf{s}^{PGA}\})$ can then be used to impose a lower
bound to the ratio $\frac{H(\{\mathbf{s}^{PGA}\})}{H(S^{\ast})}$ as follows:
\begin{equation}
L^{\prime}\triangleq L_{2}\cdot\frac{H(\{\mathbf{s}^{PGA}\})}{H(S^{G2})}%
\leq\frac{H(\{\mathbf{s}^{PGA}\})}{H(S^{\ast})}.
\label{Eq:FinalPerformanceBound}%
\end{equation}
Therefore, $L^{\prime}$ can be used as a performance bound guarantee on the
final coverage level achieved by the chosen optimal team of agents.

\paragraph*{\textbf{Characterization of optimal $t_{i}$ values given by PGA}}

We consider two agents $i$ and $j$ to be \textit{neighbors} if their sensing
regions overlap (i.e., $V(s_{i})\cap V(s_{j})\neq\emptyset$). The \textit{set
of neighbors} of agent $i$ is denoted by $B_{i}=\{j:j\neq i,V(s_{i})\cap
V(s_{j})\neq\emptyset\}$. Note that $B_{i}$ does not include $i$, therefore,
we define the \textit{closed neighborhood} of agent $i$ as $\bar{B}_{i}%
=B_{i}\cup\{i\}$. Using these neighborhood concepts, we define the following
state variable compositions to go along with $(s_{i},t_{i})$:

\begin{itemize}
\item The \textit{neighbor state} variables: $(\bar{s}_{i}^{c},\bar{t}_{i}%
^{c})$, where $\bar{s}_{i}^{c}=[\{s_{j}:j\in B_{i}\}]$ and $\bar{t}_{i}%
^{c}=[\{t_{j}:j\in B_{i}\}]$.

\item The \textit{neighborhood state} variables: $(\bar{s}_{i},\bar{t}_{i})$,
where $\bar{s}_{i}=[\{s_{j}:j\in\bar{B}_{i}\}]$ and $\bar{t}_{i}=[\{t_{j}%
:j\in\bar{B}_{i}\}]$.

\item The \textit{complementary state} variables: $(s_{i}^{c},t_{i}^{c})$,
where $s_{i}^{c}=[\{s_{j}:\forall j\neq i\}]$ and $t_{i}^{c}=[\{t_{j}:\forall
j\neq i\}]$.
\end{itemize}

Using this notation, we can now establish the following lemma.

\begin{lem}
\label{Lm:Decomposing} The objective function $H(\mathbf{s},t)$ in
\eqref{eq:coverage5} can be decomposed as,
\begin{equation}
H(\mathbf{s},t)=t_{i}H_{i}(\bar{s}_{i},\bar{t}_{i}^{c})+H_{i}^{c}(s_{i}%
^{c},t_{i}^{c}) \label{Eq:Decomposition}%
\end{equation}
where
\begin{align*}
H_{i}(\bar{s}_{i},\bar{t}_{i}^{c})  &  =\int_{V(s_{i})}R(x)\Phi_{i}%
(x)p_{i}(x,s_{i})dx-\beta\gamma_{i}, \mbox{\ and, }\\
H_{i}^{c}(s_{i}^{c},t_{i}^{c})  &  =\int_{\Omega}R(x)\left[  1-\underset
{\forall l\neq i}{\Pi}(1-\bar{p}_{l}(x,s_{l},t_{l}))\right]  dx-\beta
\underset{\forall l\neq i}{\sum}\gamma_{l}t_{l}.
\end{align*}
\end{lem}

\emph{Proof: } $H(\mathbf{s},t)$ as given in \eqref{eq:coverage5} can be
expanded as
\begin{align}
H(\mathbf{s},t)=  &  \int_{\Omega}R(x)(1-(1-\bar{p}_{i}(x,s_{i},t_{i}%
))\prod_{\forall l\neq i}(1-\bar{p}_{l}(x,s_{l},t_{l}%
)))dx\nonumber\label{Eq:DecompProofStep1}\\
&  -\beta\gamma_{i}t_{i}-\beta\sum_{\forall l\neq i}\gamma_{l}t_{l}.
\end{align}
Now, using the following relationships directly obtained from
\eqref{eq:detectionwithti}, \eqref{eq:detection}, along with the definition of
the neighbor set $B_{i}$:
\begin{align*}
\bar{p}_{i}(x,s_{i},t_{i})= &  t_{i}p_{i}(x,s_{i})\ \  &  &  \forall
x,s_{i}\in\Omega,\text{ }\forall t_{i}\in\lbrack0,1],\\
p_{i}(x,s_{i})= &  0\ \  &  &  \forall s_{i}\in\Omega,\text{ }\ x\not \in
V(s_{i}),\\
p_{i}(x,s_{i})(1-p_{j}(x,s_{j}))= &  p_{i}(x,s_{i}) &  &  \forall x\in
\Omega,\text{ }\forall j,\not \in B_{i},
\end{align*}
we can write, for all $x,s_{i},s_{l}\in\Omega$ and $t_{i},t_{l}\in\lbrack
0,1]$,
\[
\bar{p}_{i}(x,s_{i},t_{i})\prod_{\forall l\neq i}(1-\bar{p}_{l}(x,s_{l}%
,t_{l}))=t_{i}p_{i}(x,s_{i})\prod_{l\in B_{i}}(1-\bar{p}_{l}(x,s_{l},t_{l})).
\]
Using the above relationship in \eqref{Eq:DecompProofStep1}, we obtain
\eqref{Eq:Decomposition}. \hfill$\blacksquare$

Using Lemma \ref{Lm:Decomposing} we establish the following theorem which
characterizes the nature of $t_{i}^{\ast}$, the $t_{i}$ values given by the
PGA Algorithm \ref{alg:PGA}.

\begin{thm}
\label{Th:optimalt_ivalues} For any agent $i$, the values obtained from the
PGA algorithm satisfy
\begin{equation}
t_{i}^{\ast}=%
\begin{cases}
0\mbox{ when }H_{i}(\bar{s}_{i}^{\ast},\bar{t}_{i}^{c\ast})<0,\\
1\mbox{ when }H_{i}(\bar{s}_{i}^{\ast},\bar{t}_{i}^{c\ast})>0.\\
\end{cases}
\label{Eq:optimalt_ivalues}%
\end{equation}
Moreover, when $H_{i}(\bar{s}_{i}^{\ast},\bar{t}_{i}^{c\ast})=0$, the optimal
objective function value $H(\mathbf{s}^{\ast},t^{\ast})$ is invariant to
$t_{i}^{\ast}$.
\end{thm}

\emph{Proof: } Using the decomposition shown in Lemma \ref{Lm:Decomposing}, we
get
\[
\frac{\partial H(\mathbf{s},t)}{\partial t_{i}}=H_{i}(\bar{s}_{i},\bar{t}%
_{i}^{c}),
\]
where $H_{i}(\bar{s}_{i},\bar{t}_{i}^{c})$ is independent of $t_{i}$.
Therefore, when $H_{i}(\bar{s}_{i},\bar{t}_{i}^{c})\neq0$, it is clear that
the PGA cannot terminate the $t_{i}$ update process in \eqref{eq:update} until
$t_{i}$ hits a constraint boundary given by $t_{i}\in\lbrack0,1]$. The update
direction depends on the sign of $H_{i}(\bar{s}_{i},\bar{t}_{i}^{c})$ and
update process in \eqref{eq:update} will become stationary when $t_{i}$
satisfies \eqref{Eq:optimalt_ivalues}.

To prove the second statement, consider the case where $H_{i}(\bar{s}%
_{i}^{\ast},\bar{t}_{i}^{c\ast})=0$ with $t_{i}^{\ast}\in(0,1)$. Since
$H_{i}(\bar{s}_{i},\bar{t}_{i}^{c})$ is independent of $t_{i}$, if
$t_{i}^{\ast}$ is perturbed to a value $t_{i}=t_{i}^{\ast}+\Delta\in
\lbrack0,1]$, the optimality condition $H_{i}(\bar{s}_{i}^{\ast},\bar{t}%
_{i}^{c\ast})=0$ still holds true. Further, using this relationship with Lemma
\ref{Lm:Decomposing}, we can see that $H(\mathbf{s},t)$ is insensitive to a
perturbation $t_{i}^{\ast}+\Delta\in\lbrack0,1]$ when at $(\mathbf{s}%
,t)=(\mathbf{s}^{\ast},t^{\ast})$. This means that if the PGA converges to a
value $t_{i}=t_{i}^{\ast}\in(0,1)$, perturbing $t_{i}$ towards either $0$ or
$1$ will not affect the objective function value. This concludes the proof.
\hfill$\blacksquare$

\begin{rem}
\label{Rm:AdjustmentTot_izero} Using Lemma \ref{Lm:Decomposing}, it can be
further shown that, when $H_{i}(\bar{s}_{i}^{\ast},\bar{t}_{i}^{c\ast})=0$,
with $t_{i}^{\ast}\in(0,1)$, if $t_{i}^{\ast}$ is artificially perturbed, the
optimality condition for $s_{i}$ (i.e., $\frac{\partial H(\mathbf{s}%
,t)}{\partial s_{i}}=0$) still holds. However, due to such a perturbation, the
optimality conditions of neighbor agent states are affected (i.e.,
$\frac{\partial H(\mathbf{s},t)}{\partial s_{j}}\neq0,\frac{\partial
H(\mathbf{s},t)}{\partial t_{j}}\neq0,$ $j\in B_{i}$). In a such situation,
the PGA should be re-activated from the perturbed state. Also note that in
numerical simulations, occurrence of a such equivalence is unlikely.
\end{rem}

In conclusion, the proposed PGA ensures that the resulting optimal $t_{i}$
values are either $0$ or $1$. Hence, despite the relaxation of the binary
variable $t_{i}$ to $t_{i}\in\lbrack0,1]$, it provides a solution to the mixed
integer non-linear programming problem version of \eqref{eq:coverage5}, where,
for all $i$, $t_{i}$ is constrained to $t_{i}\in\{0,1\}$.

We conclude this section by observing that Lemma \ref{Lm:Decomposing} makes it
clear that in order for an agent to compute the gradients required in
\eqref{Eq:ComputedGradients} (i.e., at step 5 of Algorithm \ref{alg:PGA}), it
only needs the neighborhood state information $(\bar{s}_{i},\bar{t}_{i})$.
Therefore, in executing the PGA, agents have the capability to perform all
required computations (and subsequent actuations) in a distributed manner.

%From an implementation standpoint, this added advantage of distributed-ness reduces the required communication bandwidths and processing powers while improving the execution times and implementation complexities.

\vspace{-0.25cm}
% \newpage
\section{Numerical Results}

\label{sec:numerical} In this section, we provide several numerical results
obtained from the proposed PGA (Algorithm \ref{alg:PGA}) initialized with the
solution provided by the greedy Algorithm \ref{Alg:Greedy} discussed in
Section \ref{sec:submodularity}. The PGA method is evaluated under four
different mission space configurations named: (\romannum{1}) General,
(\romannum{2}) Room, (\romannum{3}) Maze, and, (\romannum{4}) Narrow, as shown
in Figs. \ref{fig:general}, \ref{fig:room}, \ref{fig:maze} and
\ref{fig:narrow}, respectively. The mission space is a square of size
$600\times600$ units with an event density function $R(x)$ assumed to be
uniform (i.e., $R(x)=1,\ \forall x\in F$). All simulations are initialized
with ten agents (i.e., $N=10$) and each agent's nominal sensing capacity is
selected as $p_{i0}=1$. For the use of the greedy algorithm, the ground set
$F^{D}$ is constructed by uniformly placing $100$ points in the mission space.
All reported simulation results and execution times have been obtained by
executing the algorithms on a standard desktop computer with $8.0$ GB RAM and
a $3.61$ GHz AMD eight-core processor. For convenience, we define the cost
component of the overall objective function $H(\mathbf{s},t)$ as
$C(t)=\beta\sum_{i=1}^{N}\gamma_{i}t_{i}$. Therefore, $H(s,t)=H(\mathbf{s}%
)-C(t)$ where $H(\mathbf{s})$ represents the coverage component of
$H(\mathbf{s},t)$.

\subsection{The homogeneous agent case}

\begin{table*}[!t]
\caption{Results of the proposed PGA for the homogeneous agent case.}%
\label{table:comparison_homo_one_beta}
\centering
\resizebox{.75\textwidth}{!}{
\begin{tabular}
[c]{c|ccccc|ccccc|c}\hline
Mission & \multicolumn{5}{|c|}{Initial Greedy Solution} &
\multicolumn{5}{|c|}{Final PGA Solution} & Fig.\\
Space & $N$ & $H(s)$ & $C(t)$ & $H(s,t)$ & time/s & $N$ & $H(s)$ & $C(t)$ &
$H(s,t)$ & time/s & \\\hline\hline
General & 10 & 157,111 & 142,289 & 14,822 & 2.135 & 7 & 127,225 & 99,645 &
27,580 & 1.763 & \ref{fig:general}\\
Room & 10 & 145,206 & 142,289 & 2,917 & 2.056 & 5 & 94,441 & 71,215 & 23,225 &
2.919 & \ref{fig:room}\\
Maze & 10 & 148,082 & 142,289 & 5,793 & 1.888 & 7 & 112,915 & 99,645 &
13,270 & 3.112 & \ref{fig:maze}\\
Narrow & 10 & 184,076 & 142,289 & 41,787 & 2.197 & 7 & 150,074 & 99,645 &
50,429 & 2.663 & \ref{fig:narrow}\\\hline
\end{tabular}
}
%\vspace{-0.25cm}
\end{table*}

\begin{table}[h]
\caption{Performance bound guarantees (i.e., $L^{\prime}$ in
\eqref{Eq:FinalPerformanceBound}) on the final coverage level achieved by the
optimal agent team for the homogeneous agent case.}%
\label{table:L_prime}
\centering
\begin{tabular}
[c]{c|cccc}\hline
Mission Space & $N$ & $H(S^{G2}) $ & $L_{2}$ & $L^{\prime}$\\\hline\hline
General & 7 & 114,804 & 0.651 & 0.721\\
Room & 5 & 93,086 & 0.874 & 0.886\\
Maze & 7 & 112,508 & 0.665 & 0.667\\
Narrow & 7 & 148,073 & 0.999 & 0.999\\\hline
\end{tabular}
% \vspace{-0.25cm}
\end{table}

\begin{figure}[!b]
\centering
\vspace{-0.25cm}
\begin{subfigure}[h]{\columnwidth}
\centering
\includegraphics[width=0.4\textwidth]{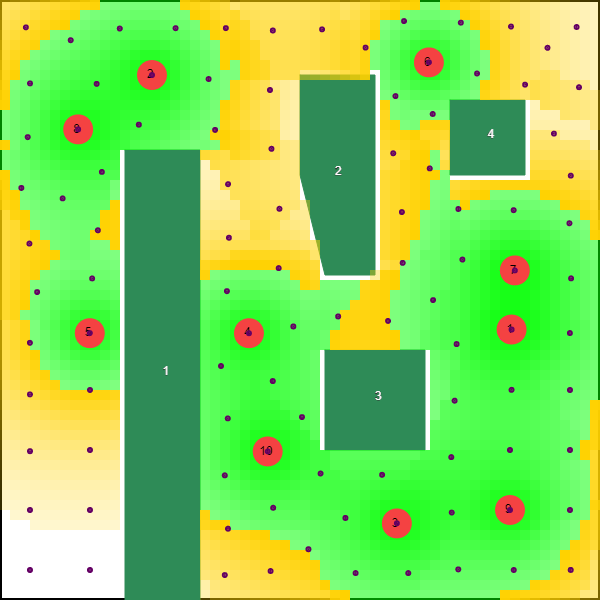}
%\hfill
\includegraphics[width=0.4\textwidth]{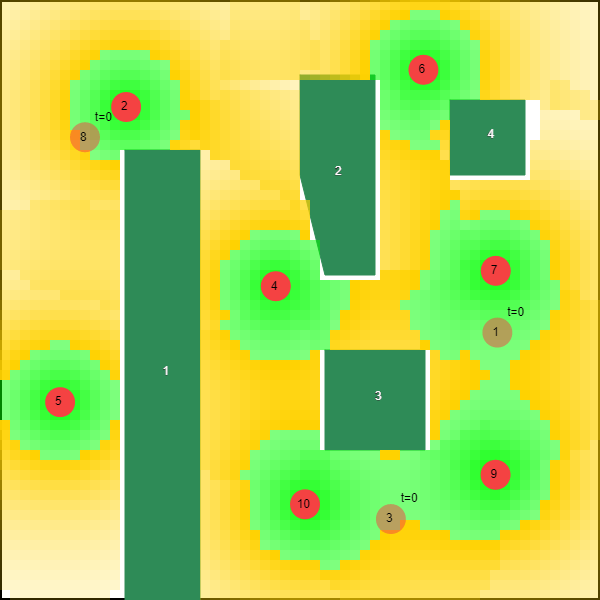}
\caption{General}
\label{fig:general}
\end{subfigure}
\vskip\baselineskip
\begin{subfigure}[h]{\columnwidth}
\centering
\includegraphics[width=0.4\textwidth]{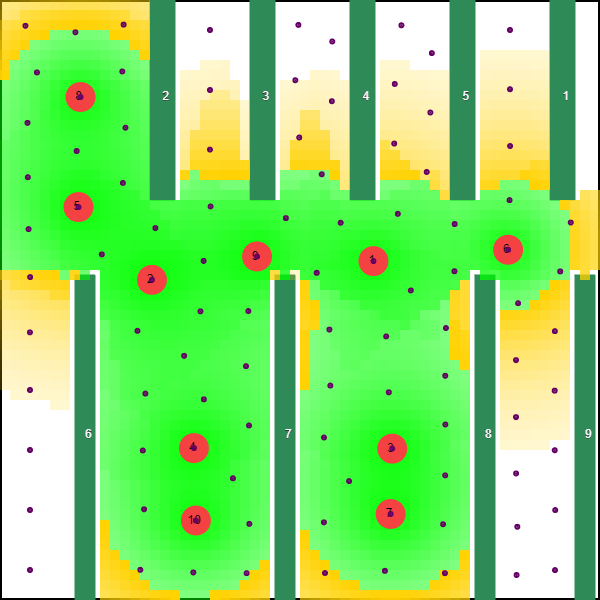}
%\hfill
\includegraphics[width=0.4\textwidth]{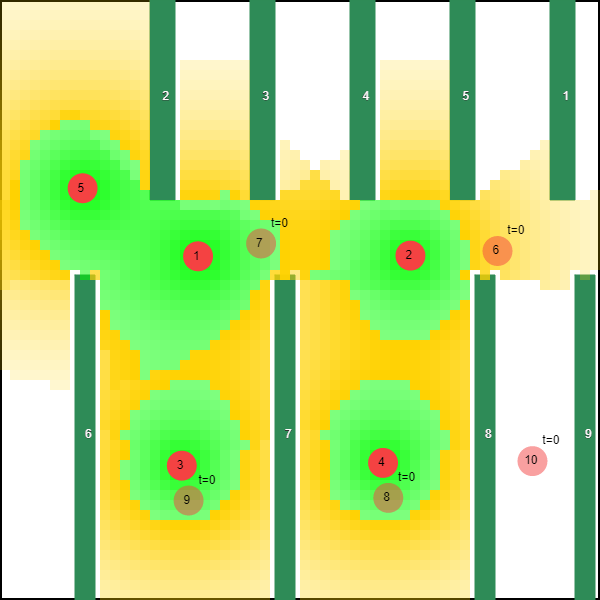}
\caption{Room}
\label{fig:room}
\end{subfigure}
\vskip\baselineskip
\begin{subfigure}[h]{\columnwidth}
\centering
\includegraphics[width=0.4\textwidth]{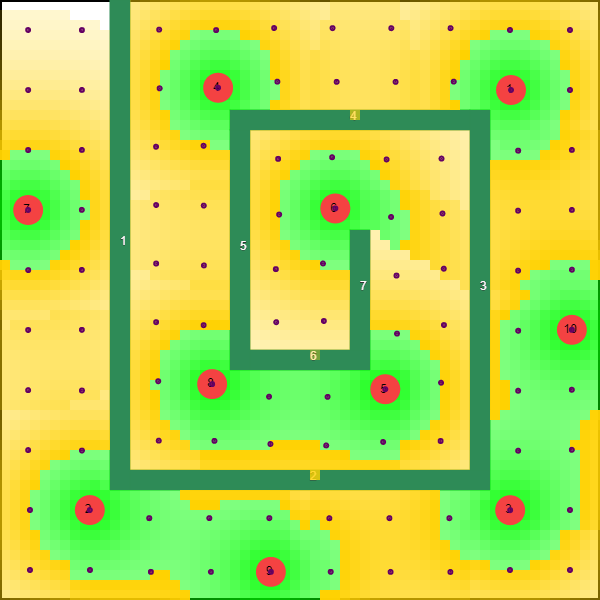}
%\hfill
\includegraphics[width=0.4\textwidth]{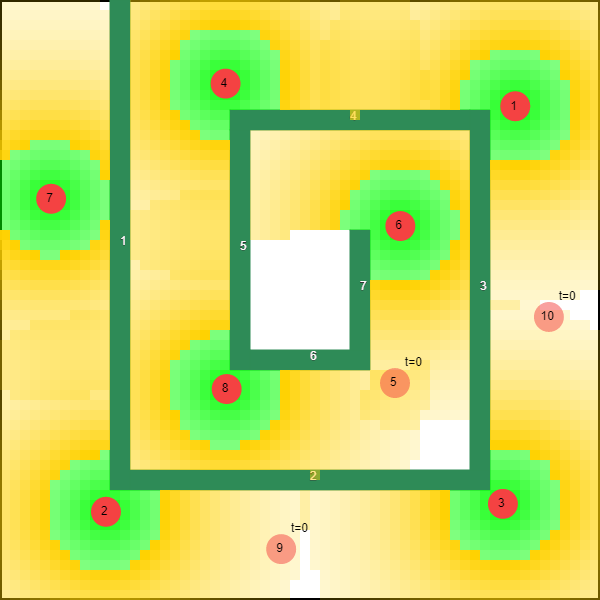}
\caption{Maze}
\label{fig:maze}
\end{subfigure}
\vskip\baselineskip
\begin{subfigure}[h]{\columnwidth}
\centering
\includegraphics[width=0.4\textwidth]{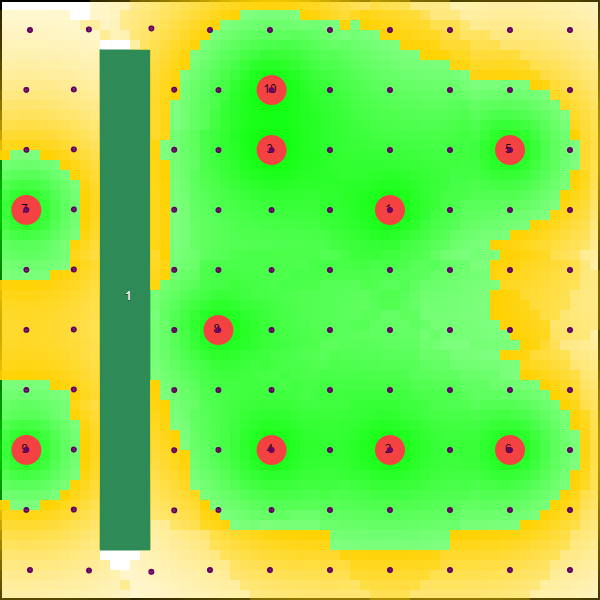}
%\hfill
\includegraphics[width=0.4\textwidth]{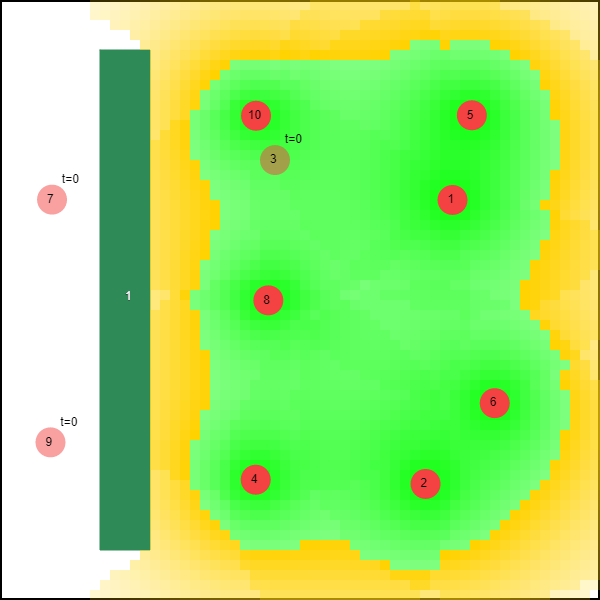}
\caption{Narrow}
\label{fig:narrow}
\end{subfigure}
\caption{Comparison of initial greedy solution (left) and projected gradient
ascent (PGA) algorithm solution (right) under different mission spaces for the
homogeneous agent case.}%
\vspace{-.25cm}
\end{figure}

We first consider the case of homogeneous agents with a sensing decay
$\lambda_{i}=0.012$, and a sensing range $\delta_{i}=200$ units in the mission
space. The weight parameters are selected as $w_{1}=0.68$ and $w_{2i}%
=1,\forall i=1,\ldots,N$ in \eqref{eq:coverage5}. The obtained results are
summarized in Tab. \ref{table:comparison_homo_one_beta} where each row
corresponds to one of the four mission space configurations defined above. The
first part of the table gives the results of the initial greedy algorithm
where the cost component is ignored. The second part gives the final results
of the PGA. Figures \ref{fig:general}, \ref{fig:room}, \ref{fig:maze} and
\ref{fig:narrow} compare the resulting system configurations at the
aforementioned two stages of the PGA. Note that the agents drawn as
light-colored disks are those with $t_{i}=0$, and, therefore, are not included
in the optimal agent team.

The overall objective value improvement over the initial greedy solution can
be seen by comparing the two $H(\mathbf{s},t)$ columns in Tab.
\ref{table:comparison_homo_one_beta}. This improvement is a result of
excluding some of the agents (from $N=10$) and obtaining solutions with
optimal agent team size $N<10$. It was observed that such agent exclusions
(i.e., $t_{i}=0$) occur when an agent's terminal location $s_{i}^{PGA}$ is in:
(\romannum{1}) A confined/narrow region where it cannot fully utilize its
sensing capabilities (e.g., see agent $10$ in Fig \ref{fig:room}), or in,
(\romannum{2}) A region which is already covered by other agents (e.g., see
agent $3$ in Fig \ref{fig:general}). As expected (see Theorem
\ref{Th:optimalt_ivalues}), all observed optimal $t_{i}$ values converged to
either $0$ or $1$ and without the need of the extra PGA step described in
Remark \ref{Rm:AdjustmentTot_izero}.

Moreover, the coverage performance bounds $L^{\prime}$ (defined in
\eqref{Eq:FinalPerformanceBound}) achieved by the optimal agent teams are
listed in Tab. \ref{table:L_prime}. From these results, we can conclude that:
(\romannum{1}) On average, the optimal team provides more than $80\%$ of the
attainable maximum coverage level, and, (\romannum{2}) In some mission spaces,
we can even guarantee near global optimality (e.g., in the Narrow mission space).

\begin{figure}[h]
\centering
\begin{subfigure}{0.49\columnwidth}
\includegraphics[width=\textwidth]{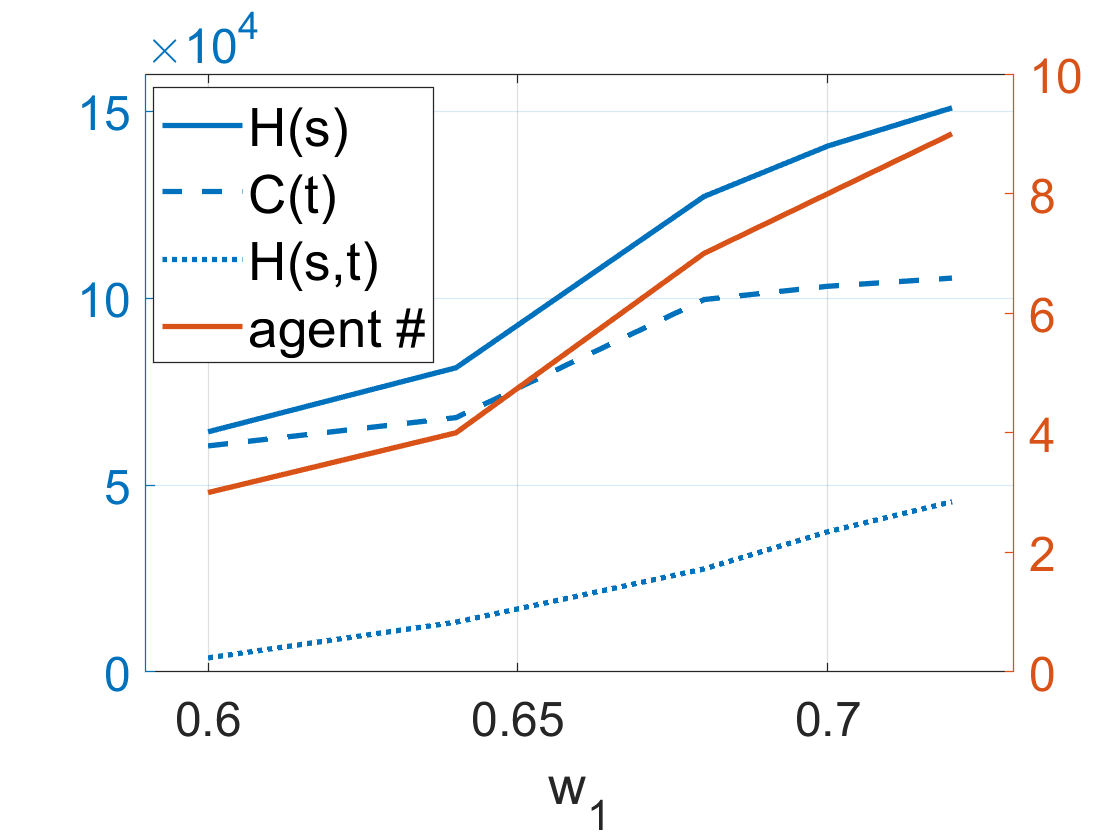}
\caption{General}
\label{f:general_Homo_varying_beta}
\end{subfigure}
\begin{subfigure}{0.49\columnwidth}
\includegraphics[width=\textwidth]{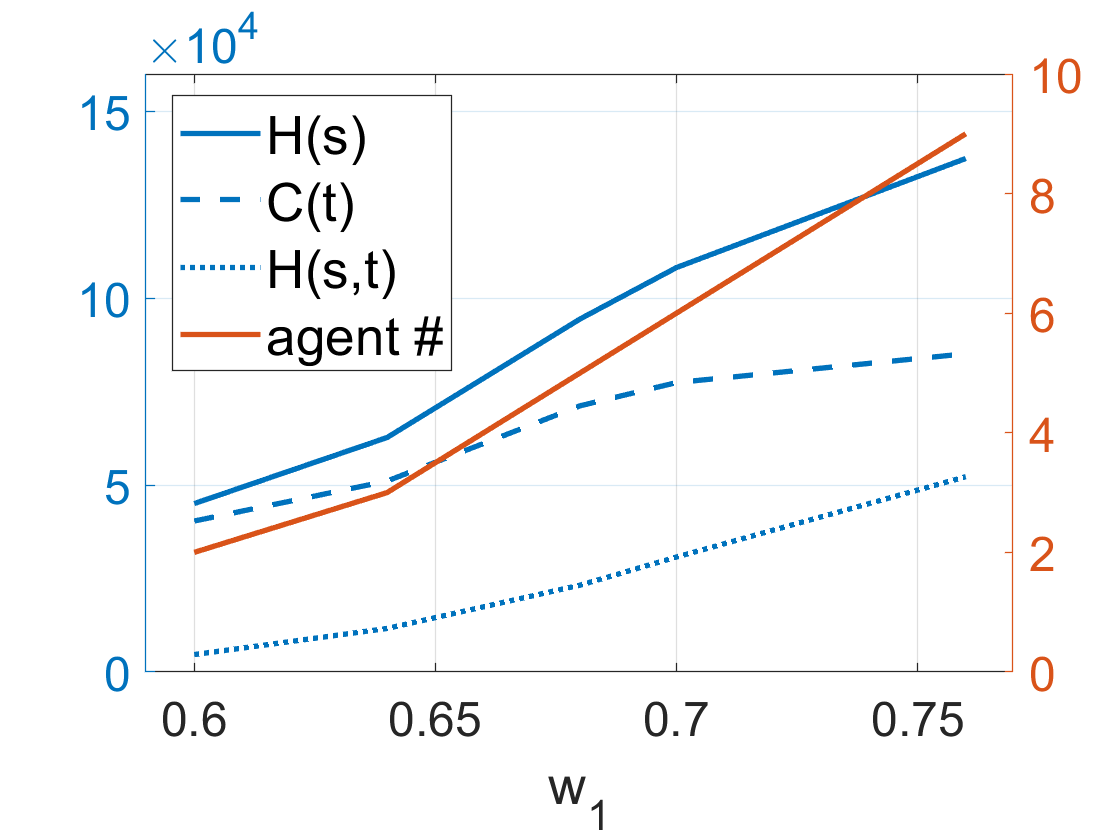}
\caption{Room}
\label{f:room_Homo_varying_beta}
\end{subfigure}
~ \begin{subfigure}{0.49\columnwidth}
\includegraphics[width=\textwidth]{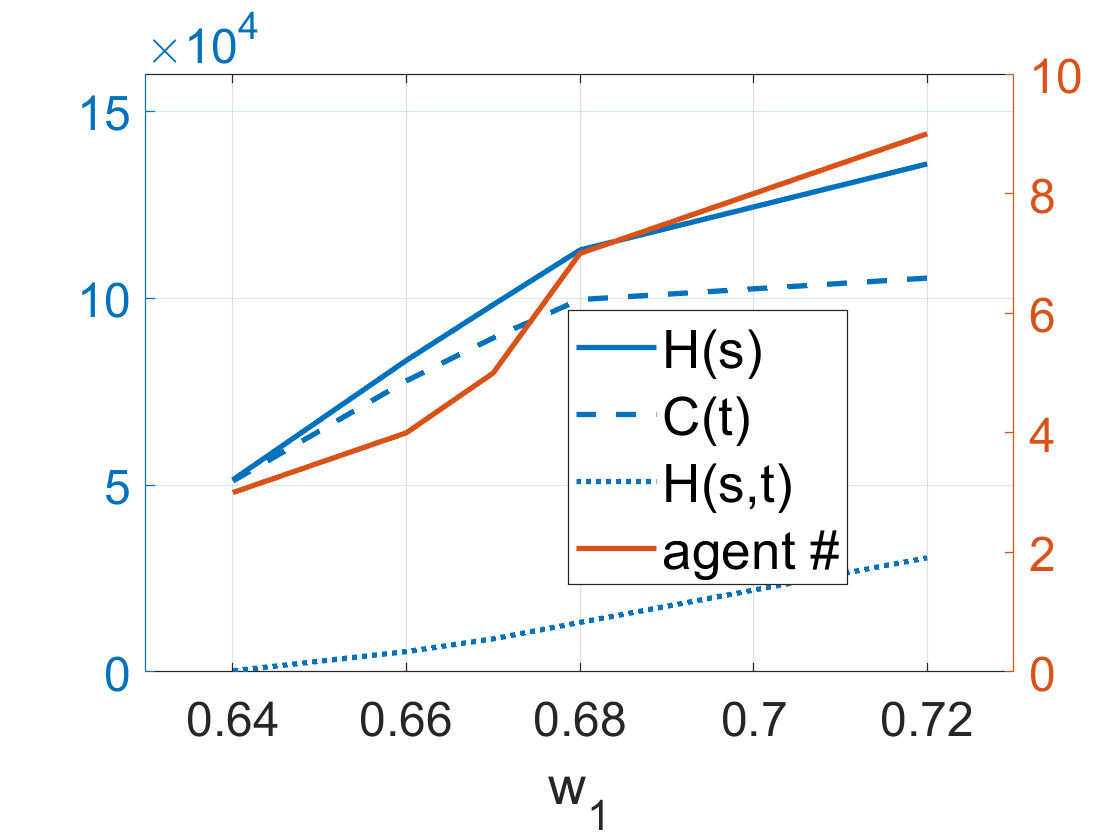}
\caption{Maze}
\label{f:maze_Homo_varying_beta}
\end{subfigure}
\begin{subfigure}{0.49\columnwidth}
\includegraphics[width=\textwidth]{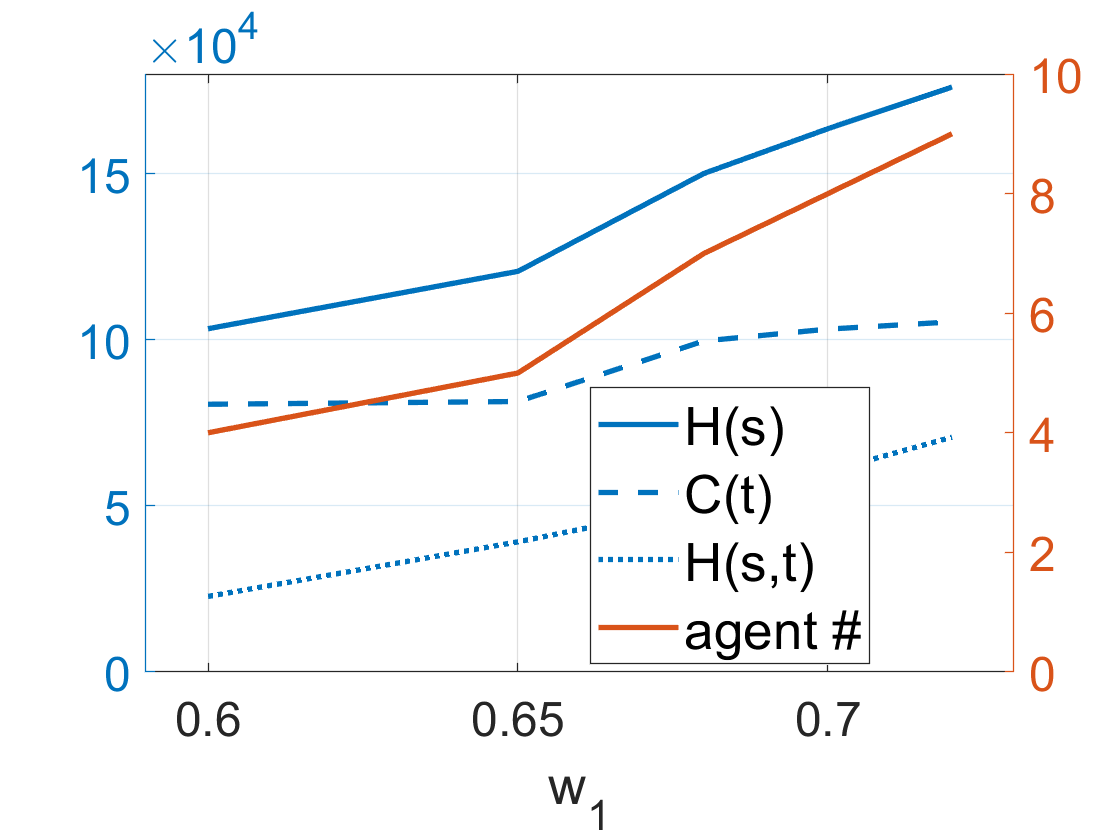}
\caption{Narrow}
\label{f:narrow_Homo_varying_beta}
\end{subfigure}
\caption{Effect of the normalization weight $w_{1}$ on the obtained PGA
solution: $H(\mathbf{s}),C(t),H(\mathbf{s},t)$ and $N$, in different mission
spaces for the homogeneous agent case.}%
\label{Fig:EffectOfBeta}%
\end{figure}

\vspace{-8mm}
\paragraph*{\textbf{The effect of the normalization factor} $\beta$}

We have studied the effect of the normalization factor $\beta$, which captures
the trade-off between the coverage performance and the team cost. Thus,
decreasing the value of $\beta$ value highlights the effect of coverage
$H(\mathbf{s})$ over the team cost $C(t)$ in the overall objective function
$H(\mathbf{s},t)$. Since, from \eqref{eq:beta_def_2}, $\beta$ directly depends
on the normalization weight $w_{1}$, we tune $w_{1}$ to get different $\beta$
values while keeping $w_{2i}=1,\ \forall i$. Figures
\ref{f:general_Homo_varying_beta}, \ref{f:room_Homo_varying_beta},
\ref{f:maze_Homo_varying_beta} and \ref{f:narrow_Homo_varying_beta} show the
effect of the normalization weight $w_{1}$ on the obtained results (i.e., the
achieved $H(\mathbf{s}),C(t),H(\mathbf{s},t)$ and $N$ values by the PGA
algorithm) for the four different mission space configurations. The main
conclusions on the behavior of the observed parameters w.r.t. $w_{1}$ are:
(\romannum{1}) It generally depends on the considered mission space,
(\romannum{2}) It is non-decreasing and piece-wise linear, and, (\romannum{3})
$H(\mathbf{s})$ grows faster than $C(t)$.

%\begin{figure}[t]
%\centering
%\begin{subfigure}{0.49\columnwidth}
%\includegraphics[width=\textwidth]{figures/general_Homo_varying_beta.png}
%\caption{General}
%\label{f:general_Homo_varying_beta}
%\end{subfigure}
%\begin{subfigure}{0.49\columnwidth}
%\includegraphics[width=\textwidth]{figures/room_Homo_varying_beta.png}
%\caption{Room}
%\label{f:room_Homo_varying_beta}
%\end{subfigure}
%~ \begin{subfigure}{0.49\columnwidth}
%\includegraphics[width=\textwidth]{figures/maze_Homo_varying_beta.png}
%\caption{Maze}
%\label{f:maze_Homo_varying_beta}
%\end{subfigure}
%\begin{subfigure}{0.49\columnwidth}
%\includegraphics[width=\textwidth]{figures/narrow_Homo_varying_beta.png}
%\caption{Narrow}
%\label{f:narrow_Homo_varying_beta}
%\end{subfigure}
%\caption{Effect of the normalization weight $w_{1}$ on the obtained PGA
%solution: $H(\mathbf{s}),C(t),H(\mathbf{s},t)$ and $N$, in different mission
%spaces for the homogeneous agent case.}%
%\label{Fig:EffectOfBeta}%
%\end{figure}

\begin{table*}[t]
\caption{Results of the proposed PGA for the heterogeneous agent case.}%
\label{table:comparison_subteams_varying_beta}
\centering
\resizebox{.75\textwidth}{!}{
\begin{tabular}
[c]{c|cccc|cccc|c}\hline
Mission & \multicolumn{4}{|c|}{Initial Greedy Solution} &
\multicolumn{4}{|c|}{Final PGA Solution} & Fig.\\
Space & $N=10$ & $H(s)$ & $C(t)$ & $H(s,t)$ & Agent Team & $H(s)$ & $C(t)$ &
$H(s,t)$ & \\\hline\hline
General & 10 & 152,272 & 177,140 & -22,868 & $\{1,2,4,5\},\{6,7,8\}$ &
124,194 & 128,174 & -3,980 & \ref{fig:general_Heterogeneous_varying_beta}\\
Room & 10 & 142,859 & 177,140 & -34,281 & $\{1,2,3\},\{7,10\}$ & 94,417 &
92,781 & 1,635 & \ref{fig:room_Heterogeneous_varying_beta}\\
Maze & 10 & 146,175 & 177,140 & -30,965 & $\{1,2,4\},\{6,7,10\}$ & 96,889 &
106,355 & -9,465 & \ref{fig:maze_Heterogeneous_varying_beta}\\
Narrow & 10 & 179,478 & 177,140 & 2,337 & $\{1,2,3,4,5\},\{6,7\}$ & 145,963 &
136,420 & 9,543 & \ref{fig:narrow_Heterogeneous_varying_beta}\\\hline
\end{tabular}
}
%\vspace{-0.25cm}
\end{table*}

\subsection{The heterogeneous agent case}

\label{SubSec:HetAgent} We now consider the heterogeneous agent case where
agents differ from each other in terms of both sensing parameters (i.e.,
sensing range $\delta_{i}$ and sensing decay $\lambda_{i}$) and cost
parameters (i.e., agent cost $\gamma_{i}$). To create such a heterogeneous
agent configuration, we first assume that the initially available $10$ agents
belong to two classes ($5$ agents per each class) as given in Tab.
\ref{table:team_setting}. Then, we set the agent cost weights to
$w_{2i}=1\ \forall i$. Based on \eqref{Eq:gamma_i_def}, under each adopted
agent class, sensing parameters $\delta_{i}$ and $\lambda_{i}$ will determine
the agent cost $\gamma_{i}$ values as shown in Tab. \ref{table:team_setting}
under the \textquotedblleft Case \ref{SubSec:HetAgent}\textquotedblright%
\ column. The normalization weight used is $w_{1}=0.58$. To make the problem
meaningful, the agent classes have been chosen so that they have complementary
sensing properties. For comparison purposes, note that in the previously
discussed homogeneous agent case, all agents belonged to Class 1.

\begin{figure}[!t]
\centering
\vspace{-0.25cm}
\begin{subfigure}[h]{\columnwidth}
\centering
\includegraphics[width=0.4\textwidth]{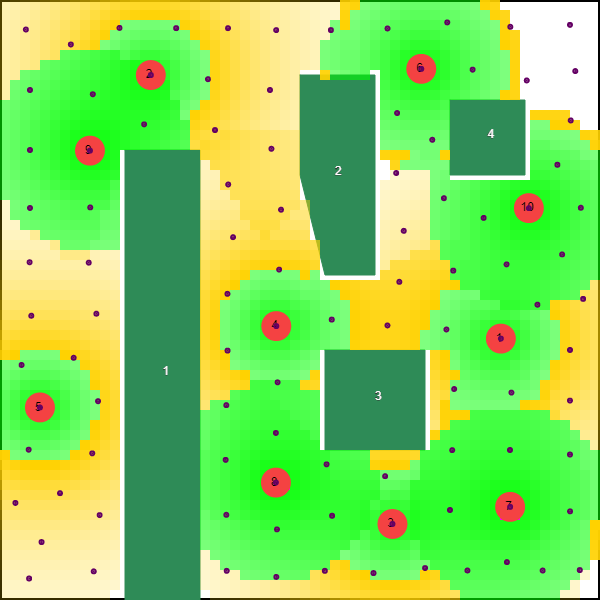}
%\hfill
\includegraphics[width=0.4\textwidth]{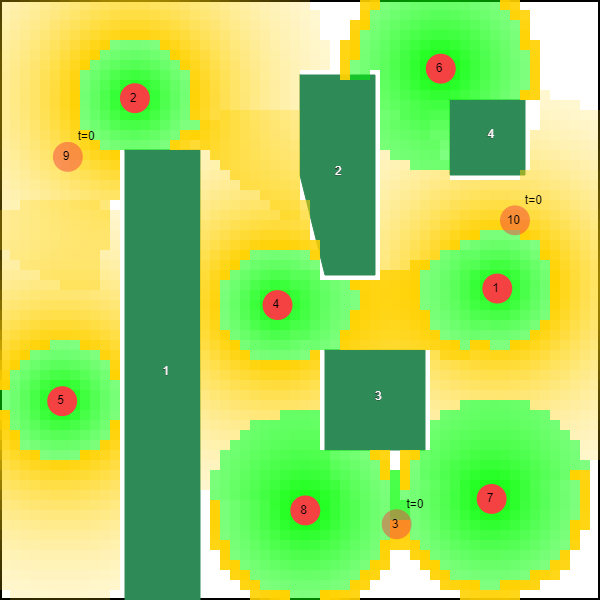}
\caption{General}
\label{fig:general_Heterogeneous_varying_beta}
\end{subfigure}
\vskip\baselineskip
\begin{subfigure}[h]{\columnwidth}
\centering
\includegraphics[width=0.4\textwidth]{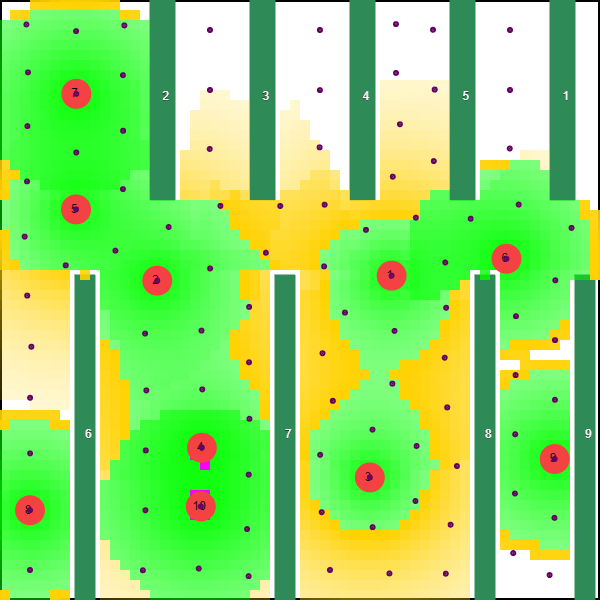}
%\hfill
\includegraphics[width=0.4\textwidth]{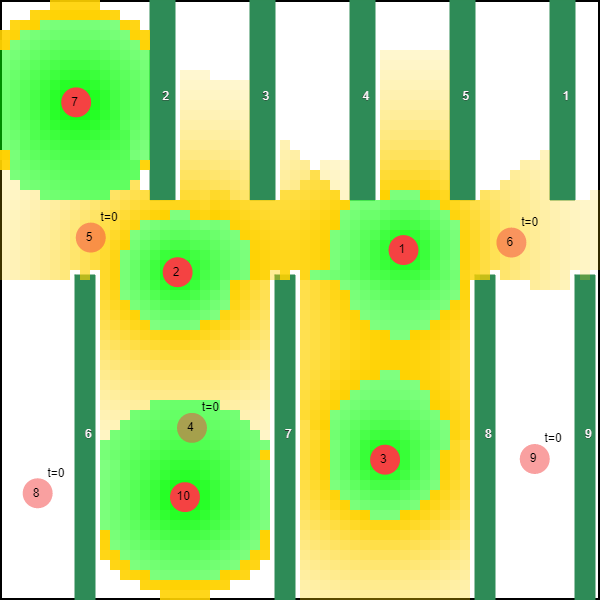}
\caption{Room}
\label{fig:room_Heterogeneous_varying_beta}
\end{subfigure}
\vskip\baselineskip
\begin{subfigure}[h]{\columnwidth}
\centering
\includegraphics[width=0.4\textwidth]{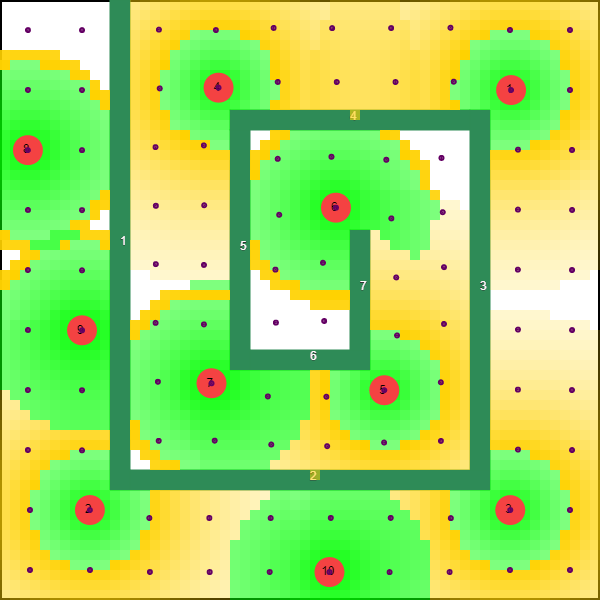}
%\hfill
\includegraphics[width=0.4\textwidth]{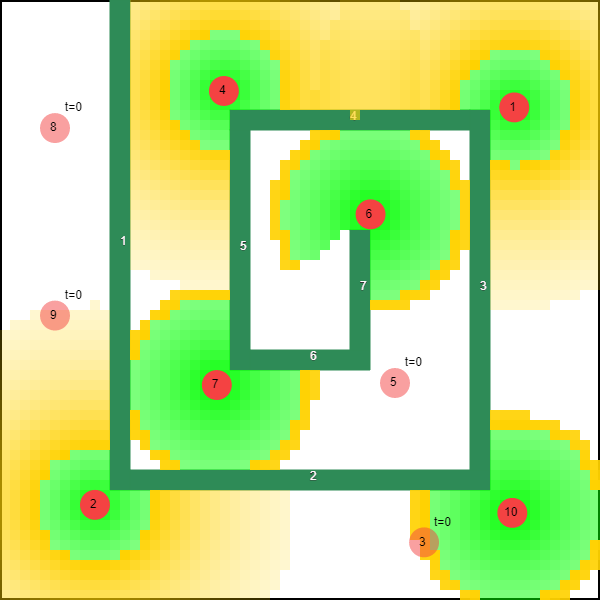}
\caption{Maze}
\label{fig:maze_Heterogeneous_varying_beta}
\end{subfigure}
\vskip\baselineskip
\begin{subfigure}[h]{\columnwidth}
\centering
\includegraphics[width=0.4\textwidth]{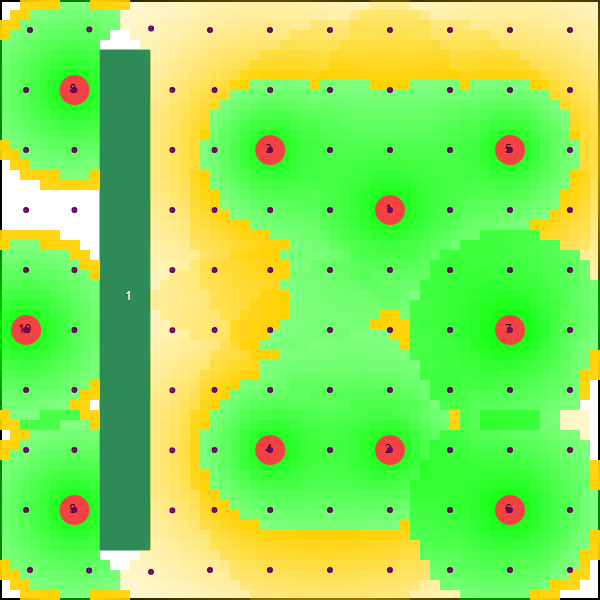}
%\hfill
\includegraphics[width=0.4\textwidth]{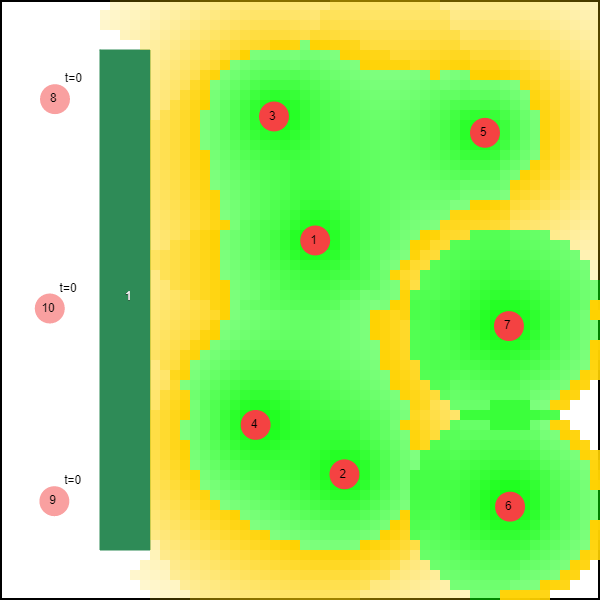}
\caption{Narrow}
\label{fig:narrow_Heterogeneous_varying_beta}
\end{subfigure}
\caption{Comparison of initial greedy solution (left) and projected gradient
ascent (PGA) algorithm solution (right) under different mission spaces for the
heterogeneous agent case.}%
\label{fig:Heterogeneous_varying_beta}%
\vspace{-0.25cm}
\end{figure}

The results obtained from the PGA algorithm are summarized in Tab.
\ref{table:comparison_subteams_varying_beta} and the corresponding optimal
agent team deployments are shown in Fig. \ref{fig:Heterogeneous_varying_beta},
both at the initial greedy step and at the final PGA solution. Similar to the
previously discussed homogeneous agent case, we can see the significant
improvement achieved in $H(\mathbf{s},t)$ by the PGA steps compared to the
initial greedy solution. It is noteworthy that the PGA algorithm has chosen
agents from both classes to form the optimal agent team. Also note that, with
the help of the initial greedy step, the PGA method has been capable of
placing agents in appropriate mission space regions well suited for their
specific sensing properties (see agent $6$ in Fig.
\ref{fig:general_Heterogeneous_varying_beta}).

The coverage performance bounds $L^{\prime}$ (defined in
\eqref{Eq:FinalPerformanceBound}) achieved by the optimal agent teams are
shown in Tab. \ref{table:comparison_hete_L_prime}. From those results, we can
conclude that, on average, the optimal agent team provides more than $75\%$ of
the attainable maximum coverage level (slightly less than the average bound
observed for the homogeneous agent case).

%\FloatBarrier

\begin{table}[h]
\caption{Different classes of agents.}%
\label{table:team_setting}%
\centering
\begin{tabular}
[c]{c|c||cc||c|c}\hline
&  & \multicolumn{2}{c||}{Sensing Para.} & Case \ref{SubSec:HetAgent} & Case
\ref{SubSec:SenWiseHetAgent}\\\hline
Class & Index & Range & Decay & $w_{2i}=1,$ & $\gamma_{i} = 30175,$\\
& $i$ & ($\delta_{i}$) & ($\lambda_{i}$) & Cost ($\gamma_{i}$) & Weight
($w_{2i}$)\\\hline\hline
1 & $1\sim5$ & 200 & 0.012 & 30175 & 1.000\\
2 & $6\sim10$ & 100 & 0.008 & 18772 & 1.607\\\hline
\end{tabular}
%\vspace{-0.25cm}
\end{table}

\begin{table*}[t]
\caption{Results of the proposed PGA for the sensing-wise heterogeneous agent
case.}%
\label{table:comparison_subteams}
\centering
\resizebox{.75\textwidth}{!}{
\begin{tabular}
[c]{c|cccc|cccc|c}\hline
Mission & \multicolumn{4}{|c|}{Initial Greedy Solution} &
\multicolumn{4}{|c|}{Final PGA Solution} & Fig.\\
Space & $N$ & $H(s)$ & $C(t)$ & $H(s,t)$ & Agent Team & $H(s)$ & $C(t)$ &
$H(s,t)$ & \\\hline\hline
General & 10 & 156,142 & 177,140 & -20,997 & $\{1,2,3,4,5\},\{\}$ & 97,398 &
88,671 & 8,726 & \ref{fig:general_Heterogeneous_one_beta}\\
Room & 10 & 145,848 & 177,140 & -31,292 & $\{1,2,3,5\},\{\}$ & 79,771 &
70,972 & 8,798 & \ref{fig:room_Heterogeneous_one_beta}\\
Maze & 10 & 146,175 & 177,140 & -30,975 & $\{1,2,3,4,5\},\{\}$ & 83,261 &
88,671 & -5,410 & \ref{fig:maze_Heterogeneous_one_beta}\\
Narrow & 10 & 179,478 & 177,140 & 2,337 & $\{1,2,3,4,5\},\{\}$ & 120,374 &
88,671 & 31,703 & \ref{fig:narrow_Heterogeneous_one_beta}\\\hline
\end{tabular}
}
%\vspace{-0.25cm}
\end{table*}

\subsection{Sensing-wise heterogeneous agent case}

\begin{table}[!b]
\caption{Performance bound guarantees (i.e., $L^{\prime}$ in
\eqref{Eq:FinalPerformanceBound}) on the final coverage level achieved by the
optimal agent team for the heterogeneous agent case.}%
\label{table:comparison_hete_L_prime}
\centering
\begin{tabular}
[c]{c|cccc}\hline
Mission & Agent Team & $H(S^{G2}) $ & $L_{2}$ & $L^{\prime}$\\
Space &  &  &  & \\\hline\hline
General & $\{2,3,4,5\},\{6,9,10\}$ & 117,923 & 0.703 & 0.740\\
Room & $\{1,2,3\},\{7,10\}$ & 86,534 & 0.853 & 0.931\\
Maze & $\{1,2,4\},\{6,7,10\}$ & 91,203 & 0.703 & 0.747\\
Narrow & $\{1,2,3,4,5\},\{6,7\}$ & 144,852 & 0.651 & 0.656\\\hline
\end{tabular}
%\vspace{-0.25cm}
\end{table}

Our purpose here is to highlight the importance of having different agent
costs $\gamma_{i}$ when the sensing parameters of the agents are different. We
also highlight the importance of using the sensing capability (i.e.,
$\kappa_{i}$) dependent agent costs as proposed in \eqref{Eq:gamma_i_def}.
Unlike the previously discussed heterogeneous agent case, here we use a fixed
agent cost $\gamma_{i}=30175$ across all agent classes. To achieve this under
\eqref{Eq:gamma_i_def}, we manipulate the agent cost weight $w_{2i}$
parameters in each agent class, as given in Tab. \ref{table:team_setting}
column \textquotedblleft Case \ref{SubSec:SenWiseHetAgent}\textquotedblright.
As a result of this manipulation, despite the differences in sensing
parameters over different agents, the agent costs $\gamma_{i}$ across all
agents become identical. The normalization weight used is $w_{1}=0.58$.

\begin{figure}[t]
\centering
\begin{subfigure}{0.4\columnwidth}
\includegraphics[width=\textwidth]{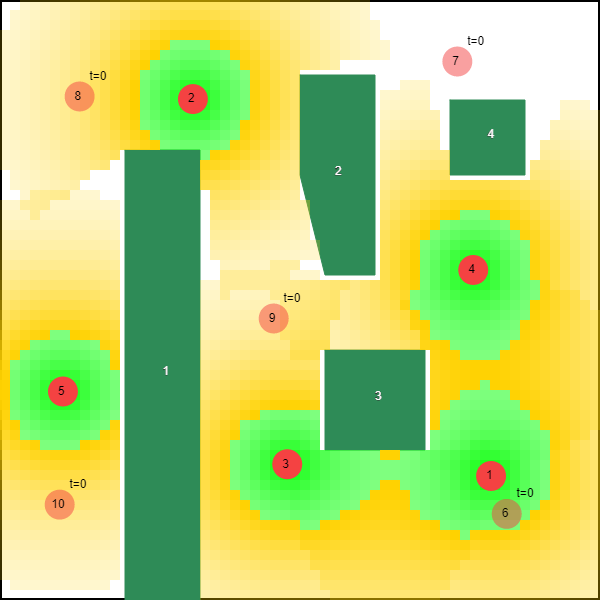}
\caption{General}
\label{fig:general_Heterogeneous_one_beta}
\end{subfigure}
\begin{subfigure}{0.4\columnwidth}
\includegraphics[width=\textwidth]{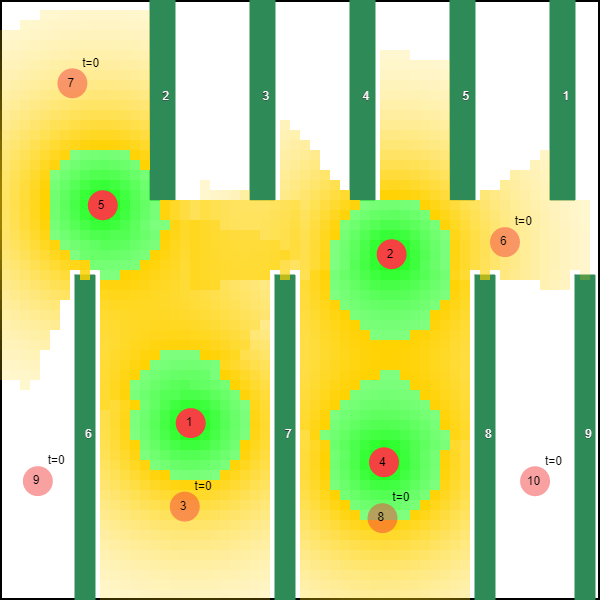}
\caption{Room}
\label{fig:room_Heterogeneous_one_beta}
\end{subfigure}
~ \begin{subfigure}{0.4\columnwidth}
\includegraphics[width=\textwidth]{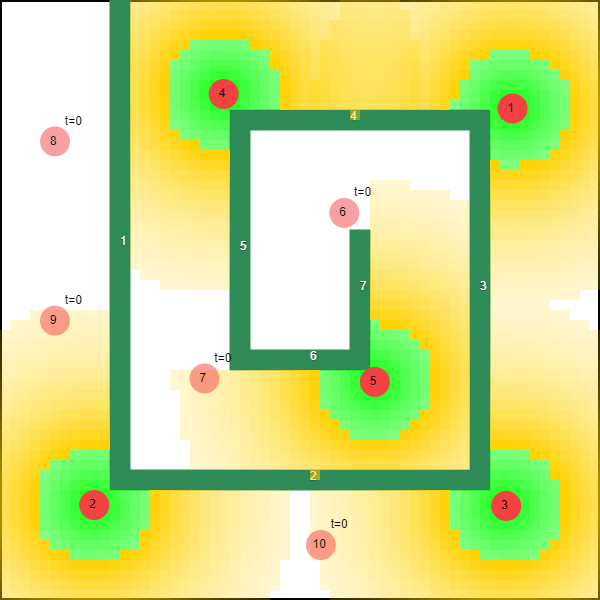}
\caption{Maze}
\label{fig:maze_Heterogeneous_one_beta}
\end{subfigure}
\begin{subfigure}{0.4\columnwidth}
\includegraphics[width=\textwidth]{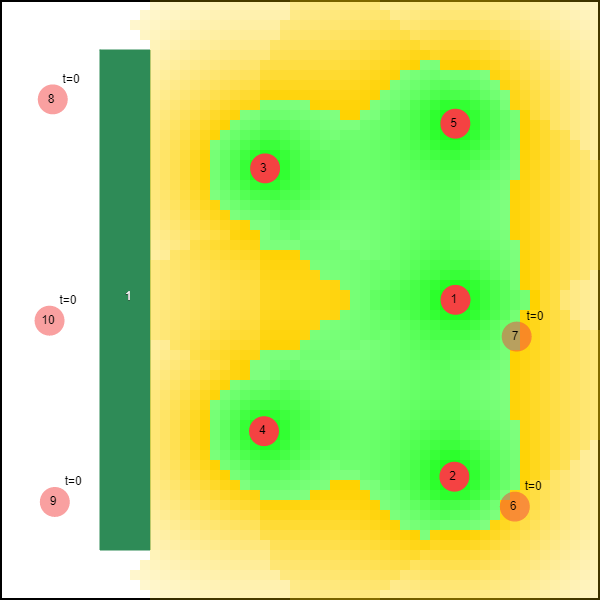}
\caption{Narrow}
\label{fig:narrow_Heterogeneous_one_beta}
\end{subfigure}
\caption{The obtained final PGA solution under different mission spaces for
the sensing-wise heterogeneous agent case.}%
\label{fig:Heterogeneous_one_beta}%
\end{figure}

Since all the other problem settings are identical to the previously discussed
heterogeneous agent case (in subsection \ref{SubSec:HetAgent}), the initial
greedy step of the PGA algorithm will yield the same agent deployment.
However, the associated total agent cost $C(t)$ will be different due to the
modification of agent cost parameters $w_{2i}$ and $\gamma_{i}$ compared to
that of the heterogeneous agent case. The numerical results obtained are
summarized in Tab. \ref{table:comparison_subteams} and the optimal agent team
deployments are shown in Fig. \ref{fig:Heterogeneous_one_beta}. The coverage
performance bounds $L^{\prime}$ (defined in \eqref{Eq:FinalPerformanceBound})
achieved by the optimal agent team are tabulated in Tab.
\ref{table:hete_same_gamma_L_prime}.

As expected, when identical agent costs are used despite their differences in
sensing capabilities, the resulting PGA solution gives preference to agents
with higher sensing capabilities. As a result, the optimal agent team is
inherently biased towards Class 1 agents (see Tab.
\ref{table:comparison_subteams} and notice $\kappa_{1}>\kappa_{2}$ due to the
$\delta_{i},\lambda_{i}$ values $i=1,2$). Clearly, in real-world applications
one expects more capable sensors to have higher costs.

\label{SubSec:SenWiseHetAgent}

\begin{table}[t]
\caption{Performance bound guarantees (i.e., $L^{\prime}$ in
\eqref{Eq:FinalPerformanceBound}) on the final coverage level achieved by the
optimal agent team for the sensing-wise heterogeneous agent case.}%
\label{table:hete_same_gamma_L_prime}
\centering
\begin{tabular}
[c]{c|cccc}\hline
Mission Space & Agent Team & $H(S^{G2}) $ & $L_{2}$ & $L^{\prime}%
$\\\hline\hline
General & $\{1,2,3,4,5\},\{\}$ & 95,633 & 0.729 & 0.742\\
Room & $\{1,2,3,5\},\{\}$ & 73,864 & 0.813 & 0.878\\
Maze & $\{1,2,3,4,5\},\{\}$ & 82,957 & 0.703 & 0.706\\
Narrow & $\{1,2,3,4,5\},\{\}$ & 117,231 & 0.651 & 0.668\\\hline
\end{tabular}
%\vspace{-0.25cm}
\end{table}

\subsection{Comparison with a commercial optimization solver}

In comparing the solutions given by the proposed PGA method to those of a
commercially available optimization problem solver, there are two constraining
factors to consider: (\romannum{1}) The coverage component of the objective
function in \eqref{eq:coverage5} is non-convex, non-linear, and discontinuous.
As a result, even though the original version of \eqref{eq:coverage5} is a
mixed-integer non-linear program (MINLP) (where $t_{i}\in\{0,1\},\ \forall
i$), we were constrained to using a generic non-linear program (NLP) solver.
Therefore, in order to find the optimal binary decision variables (i.e., $t$),
we applied the NLP solver exhaustively over all possible integer values (we
refer to this as the \textquotedblleft brute force\textquotedblright\ method).
(\romannum{2}) When obstacles are present in the mission space, the feasible
space for each agent becomes non-convex (in our case, this complicates the
objective function as well). Since representing such constraints and feeding
them to a generic optimization problem solver is difficult, we confine our
study to an obstacle-less (blank) mission space.

\begin{figure}[h]
\centering
\begin{subfigure}{0.31\columnwidth}
\includegraphics[width=\textwidth]{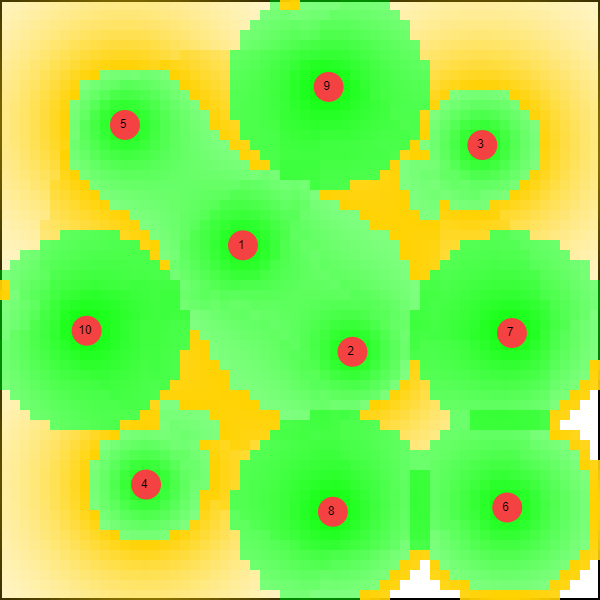}
\caption{PGA ($w_1=1$): \\ $H(\mathbf{s})=199037$ \\ Ex.T. $= 2.699 s$}
\label{Fig:NumericalComparisonPGA}
\end{subfigure}
\hfill\begin{subfigure}{0.31\columnwidth}
\includegraphics[width=\textwidth]{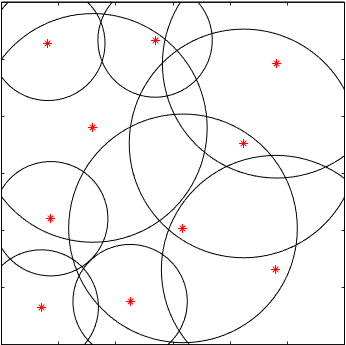}
\caption{BF\textsubscript{1} (1 Iter.): \\ $H(\mathbf{s}) = 198083$ \\ Ex.T. $= 246.523 s$}
\label{Fig:NumericalComparisonBF_1}
\end{subfigure}
\hfill\begin{subfigure}{0.31\columnwidth}
\includegraphics[width=\textwidth]{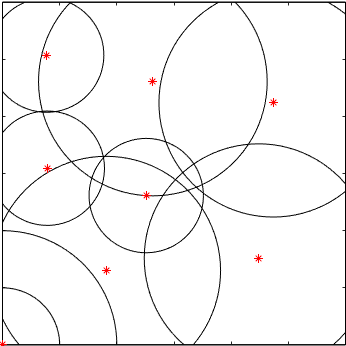}
\caption{BF\textsubscript{2} (1 Iter.): \\ $H(\mathbf{s}) = 164345$ \\ Ex.T. $= 728.374$}
\label{Fig:NumericalComparisonBF_2}
\end{subfigure}
\caption{Optimal agent configurations, coverage levels, and execution times
obtained for the multi-agent coverage problem (see \eqref{eq:coverage1}) with
$10$ heterogeneous agents (see Tab. \ref{table:team_setting}) in a blank
mission space using (a) PGA algorithm, (b) Brute force method 1
(BF\textsubscript{1}), and, (c) Brute force method 2 (BF\textsubscript{2}).}%
\label{Fig:NumericalComparison}%
\end{figure}

The NLP solver used is the \textit{interior point method} implemented under
the \textit{fmincon} command in MATLAB\textsuperscript{\textregistered}. The
available agents and their sensing capabilities are given in Tab.
\ref{table:team_setting}. In the brute force approach, each iteration
considers a specific agent team and computes the optimal coverage solution.
Two brute force methods (BF\textsubscript{1} and BF\textsubscript{2}) were
used depending on the agent initialization in order to highlight the effect of
such initialization. Specifically, in BF\textsubscript{1}, agents are
initialized randomly and in BF\textsubscript{2}, agents are initialized in a
corner of the mission space such that the $l$\textsuperscript{th} agent
($\forall l$) is placed at $s_{l}=(5+5l,5+5l)$. Note that when the
normalization weight is $w_{1}=1$ (see \eqref{eq:beta_def_2},
\eqref{eq:coverage5}), the PGA method basically solves the optimal coverage
problem. This enables a direct comparison of the performance of the PGA method
(when $w_{1}=1$) with that of single iterations of BF\textsubscript{1} and
BF\textsubscript{2}. This comparison is shown in Fig.
\ref{Fig:NumericalComparison} and it confirms that the proposed PGA method:
(\romannum{1}) Delivers better coverage levels, and, (\romannum{2}) Shows
extremely low execution times compared to BF\textsubscript{1} or
BF\textsubscript{2}. Another conclusion is that the random initialization has
helped the BF\textsubscript{1} method to achieve better coverage and execution
times compared to that of BF\textsubscript{2}.

\begin{figure}[h]
\centering
\begin{subfigure}{0.49\columnwidth}
\includegraphics[width=\textwidth]{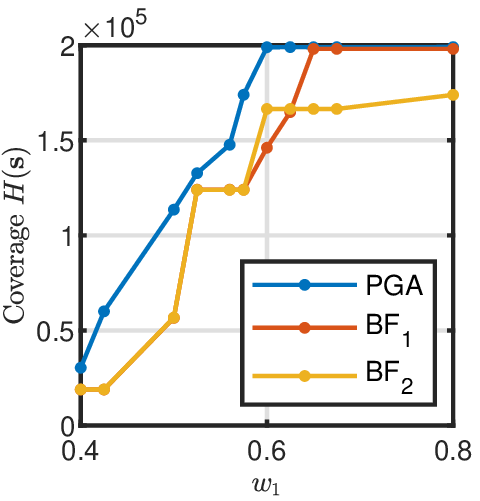}
\caption{Coverage performance: $H(s)$}
\label{Fig:NumericalComparison2Coverage}
\end{subfigure}
\begin{subfigure}{0.49\columnwidth}
\includegraphics[width=\textwidth]{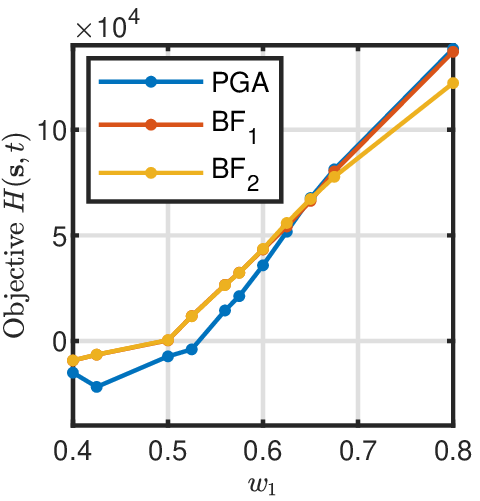}
\caption{Overall performance: $H(\mathbf{s},t)$}
\label{Fig:NumericalComparison2Objective}
\end{subfigure}
\caption{Comparison of coverage performance $H(\mathbf{s})$ and
overall performance $H(\mathbf{s},t)$ for different normalization weights
$w_{1}$ in \eqref{eq:beta_def_2}.}%
\label{Fig:NumericalComparison2}%
\end{figure}

Under the information in Tab \ref{table:team_setting}, there are $35$ possible
agent team configurations. Therefore, $35$ brute force iterations were
required to determine the optimal agent configuration. As the next step, the
agent cost related parameters $\beta$ and $\gamma_{i}$ were computed using the
prespecified weights $w_{1}$ and $w_{2i}$. Then, the best agent team
composition, which maximizes the overall objective $H(\mathbf{s},t)$, is
identified from simply searching through the previously generated results. A
comparison of the obtained results in terms of the coverage $H(\mathbf{s})$
and the overall objective $H(\mathbf{s},t)$ when the weight $w_{1}$ is varied
is shown in Fig. \ref{Fig:NumericalComparison2}. The average value of the
execution times observed in each method is given in Tab.
\ref{Tab:ExecutionTimes}.

\begin{table}[h]
\caption{Observed average execution times.}%
\label{Tab:ExecutionTimes}%
\centering
\begin{tabular}
[c]{|c|c|c|c|}\hline
Method & PGA & BF\textsubscript{1} & BF\textsubscript{2}\\\hline
Average execution time / ($s$) & 4.56 & 4328.13 & 8845.83\\\hline
\end{tabular}
\end{table}

Our main conclusions from this comparison are: (\romannum{1}) The PGA method
delivers better coverage levels $H(\mathbf{s})$ across all $w_{1}$ values
used, and, (\romannum{2}) As $w_{1}$ increases, the PGA method performs better
than brute force methods in terms of $H(\mathbf{s},t)$, and, most importantly,
(\romannum{3}) The average execution time required for the PGA method is
extremely low compared to brute force approaches (by a factor of $10^{-3}$).
Finally, we also emphasize the scalability that the PGA method offers due to
its distributed nature.

%\vspace{-0.25cm}
%\vspace{-30cm}

% \begin{figure}[h]
% \centering
% \begin{subfigure}{0.49\columnwidth}
% \includegraphics[width=\textwidth]{figures/FullyHeterogeneousCoverage.eps}
% \caption{Comparative performance\\ \centering of $H(s)$}
% \label{Fig:NumericalComparison2Coverage}
% \end{subfigure}
% \begin{subfigure}{0.49\columnwidth}
% \includegraphics[width=\textwidth]{figures/FullyHeterogeneousObjective.eps}
% \caption{Comparative performance\\ \centering of  $H(\mathbf{s},t)$}
% \label{Fig:NumericalComparison2Objective}
% \end{subfigure}
% \caption{Comparative performance of coverage objective $H(\mathbf{s})$ and
% overall objective $H(\mathbf{s},t)$ over different normalization weights
% $w_{1}$ in \eqref{eq:beta_def_2}.}%
% \label{Fig:NumericalComparison2}%
% \end{figure}

\section{Conclusions}

\label{sec:conclusions} Multi-agent coverage problem is well-studied when a
fixed number of homogeneous agents is to be deployed. In contrast, we address
the multi-agent coverage problem where the number of agents to be used is
flexible and the available agents are both heterogeneous and have an
associated cost value. We have addressed this optimal agent team composition
problem by constructing an objective function combining the overall agent team
cost with the coverage level delivered by the agent team. An $l_{1}$
regularizer is introduced to transform the agent team composition problem into
a resource (sensing capacity) allocation problem with no extra non-convexity
present. This problem is then solved using a projected gradient ascent (PGA)
algorithm initialized through a greedy algorithm and shown to recover the
integer-valued variables that were originally relaxed. Further, based on
submodularity theory, we have derived tighter performance bounds showing that
the PGA algorithm can often lead to near-global-optimal solutions. The
effectiveness of the PGA algorithm in diverse mission spaces and heterogeneous
multi-agent scenarios has been validated. Additionally, a comparison study
with results obtained from a commercial MINLP solver show the efficiency of
our proposed PGA method. An interesting future research direction would be to
investigate the applicability of the proposed approach to other multi-agent
problems with heterogeneous agents.

%\textcolor{red}{
%The advantages of the proposed algorithm are considering the combinatorial heterogeneous team composition in the multi-agent coverage problem and proposing tighter bounds of the greedy algorithm than those in the literature. A possible direction for future research is to overcome the non-convexity in a online manner.
%Even though the proposed PGA algorithm have been formulated considering a two-dimensional mission space, it is applicable for any $N$-dimensional mission space, upon appropriately modifying the representations of the agent sensing models and the obstacles.}

{
The key advantages of the proposed PGA algorithm are: (\romannum{1}) it is capable of solving the combinatorial
problem of determining the optimal agent team composition, (\romannum{2}) it addresses a number of challenges
raised due to agent heterogeneity, and (\romannum{3}) it is characterized by tighter performance bounds for the obtained final
solution. Finally we point out that even though the proposed PGA algorithm has been formulated considering a
two-dimensional mission space, it is applicable to any $N$-dimensional mission space upon appropriately modifying
the representations of the agent sensing models and the obstacles.
}

%%%%%%%%%%%%%%%%%%%%%%%%%%%%%%%%%%%%%%%%%%%%%%%%%%%%%%%%%%%%%%%%%%%%%%%%%%%%%%%%

%ifCLASSOPTIONcaptionsoff
%\newpage\fi

%\bibliographystyle{ieeeconf}
%\bibliographystyle{ieeetr}
%\bibliographystyle{siam}
\bibliographystyle{abbrv}
\bibliography{reflib}
\balance

\end{document}